\documentclass[traditabstract]{aa} % for the abstract without structuration 

\usepackage{subfig}
\usepackage{float}
\usepackage{natbib}
\usepackage{lscape}
\usepackage{rotating}
\usepackage{graphicx}
\usepackage{amsmath}
\usepackage{enumerate}
\usepackage{color}
\usepackage{afterpage}
\usepackage[varg]{txfonts}

\begin{document}

  \title{Models of low-mass helium white dwarfs including gravitational settling, thermal and chemical diffusion, and rotational mixing}
   \subtitle{}

   \author{A. G. Istrate \inst{1,2}\fnmsep\thanks{e-mail: istrate@uwm.edu},
          P. Marchant \inst{1},           
          T. M. Tauris\inst{3,1}, 
          N. Langer\inst{1}, 
          R. J. Stancliffe\inst{1}
          \and
          L. Grassitelli\inst{1}
           }

   \authorrunning{Istrate et al.}
   \titlerunning{A new grid of ELM WD models}

   \institute{Argelander-Institute f\"ur Astronomie, Universit\"at Bonn, Auf
              dem H\"ugel 71, 53121 Bonn, Germany
               \and 
              Center for Gravitation, Cosmology and Astrophysics,
               Department of Physics, University of Wisconsin-Milwaukee,
               P.O. Box 413, Milwaukee, WI 53201, USA             
              \and
              Max-Planck-Institut f\"ur Radioastronomie, 
              Auf dem H\"ugel 69, 53121 Bonn, Germany 
              }

\date{Received May, 2016, Accepted June, 2016}

\abstract{A large number of extremely low-mass helium white dwarfs (ELM~WDs) have been discovered in recent years. The majority of them are found in close binary systems suggesting they are 
formed either through a common-envelope phase or via stable mass transfer in a low-mass X-ray binary (LMXB) or a cataclysmic variable (CV) system.
Here, we investigate the formation of these objects through the LMXB channel with emphasis on the proto-WD evolution in environments with different metallicities. 
We study for the first time the combined effects of rotational mixing and element diffusion (e.g. gravitational settling, thermal  and chemical diffusion) on the evolution of proto-WDs and on the cooling properties of the resulting WDs.
We present state-of-the-art binary stellar evolution models computed with MESA for metallicities of $Z=0.02$, 0.01, 0.001 and 0.0002, producing WDs with masses between $\sim 0.16-0.45\;M_{\odot}$.  
Our results confirm that element diffusion plays a significant role in the evolution of proto-WDs that experience hydrogen shell flashes. The occurrence of these flashes  produces a clear dichotomy in the cooling timescales of ELM~WDs, which has important consequences e.g. for the age determination of binary millisecond pulsars. In addition, we confirm that the threshold mass at which this dichotomy occurs depends on metallicity. Rotational mixing is found to counteract the effect of gravitational settling in the surface layers of young, bloated ELM~proto-WDs and therefore plays a key role in determining their surface chemical abundances, i.e. the observed presence of metals in their atmospheres. We predict that these proto-WDs have helium-rich envelopes through  a significant part of their lifetime. This is of great importance as helium  is a crucial ingredient in the driving  of the $\kappa-$mechanism suggested for the newly observed ELM proto-WD pulsators.
 However, we  find that  the number of hydrogen shell flashes and, as a result, the hydrogen envelope mass at the beginning of the cooling track, are not influenced significantly by rotational mixing.  
In addition to being dependent on proto-WD mass and metallicity, the hydrogen envelope mass of the newly formed proto-WDs depends on whether or not the donor star experiences a temporary contraction when the H-burning shell crosses the hydrogen discontinuity left behind by the convective envelope.
The hydrogen envelope at detachment, although small compared to the total mass of the WD,  contains enough angular momentum such that the spin frequency of the resulting WD on the cooling track is  well above the orbital frequency.}

\keywords{white dwarfs --- stars: low-mass --- stars: evolution --- binaries: close --- binaries: X-rays --- pulsars: general}

\maketitle

%%%%%%%%%%%%%%%%%%%%%%%%%%%%%%%%%%%%%%%%%%%%%%%%%%

%%%%%%%%%%%%%%%%% BODY OF PAPER %%%%%%%%%%%%%%%%%%
\section{Introduction}\label{section:introduction}
Extremely low-mass white dwarfs (ELM~WDs) are low-mass helium-core WDs with masses below $0.2-0.3\;M_{\odot}$ and with surface gravities of $5<\log g<7$ \citep{elm5}.  
A large number of such objects have been discovered in recent years through dedicated or general surveys such as ELM, SPY, WASP, SDSS and the \textit{Kepler} mission \citep[e.g.][]{elm1,elm2,elm3,elm4,elm5, koester09,maxted11,kepler15, warren2016}. Soon after the discovery of the first ELM WDs, it was recognised that they have to be a product of binary evolution \citep{marsh95}.
From an evolutionary point of view, these ELM~WDs cannot be formed from single-star progenitors as the nuclear evolution timescale of such low-mass objects would exceed the Hubble time -- unless they have an extremely high metallicity \citep{kilic07} or the star lost its envelope from an inspiralling giant planet \citep{nt98}. Indeed, the vast majority of ELM~WDs are found in binary systems with a companion star such as a neutron star in millisecond pulsar (MSP) systems \citep{kerkwijk05}, 
an A-type star in EL~CVn-type systems \citep{maxted14} or another (typically a carbon-oxygen) WD. 
ELM~WDs have been discovered in various environments, from the Galactic disk to open and globular clusters \citep{rivera15,cadelano15}, and thus they can be formed from progenitors with different metallicities. 
   
The revived interest in ELM~WDs was fostered by the discovery of pulsations in several of these objects \citep{hermes12,hermes13b,hermes13,kilic15} as well as ELM proto-WDs \citep{maxted2013, maxted2014, corti2016, gianninas2016}. The ELM~WD pulsators extend the ZZ Ceti instability strip to lower effective temperatures and higher luminosities. This instability strip contains stars with a convective driving 
mechanism for pulsations acting at the base of the convective zone associated with hydrogen recombination \cite[e.g.][]{grootel2013}. In the newly discovered ELM~proto-WD pulsators, the excitation mechanism is instead the usual $\kappa-$ mechanism for which the presence of He in the envelope is thought to play a key role \citep{jeffsaio2013,corsico16}. The pulsational behaviour of ELM~WDs and ELM proto-WDs provide an unique  insight into their interior properties, such as the hydrogen envelope mass and their total mass and rotation rate,  which will place stronger constraints on the theoretical models \citep[e.g.][]{corsico14,corsico14b,corsico16}.  
       
Another interesting and not completely understood feature of ELM~WDs is the observed presence of metals in their atmospheres. \cite{gianninas14} provided for the first time systematic measurements of the atmospheric abundances of He, Ca and Mg for this type of stars and examined their distribution as a function of effective temperature and mass. In the observed sample, all the WDs with $\log~g<5.9$ show Ca~II~K lines, suggesting that the presence of metals in these objects is a ubiquitous phenomenon, possibly linked to their evolution. Detailed abundance analyses exist for only a handful of objects \citep{kaplan13,gianninas14b,hermes14b,latour15} but already suggest a diversity of metallicities, as in the case of sdB stars. Gravitational settling depletes the metals in the atmospheres of WDs on a very short timescale compared to their evolutionary timescale \citep{vauclair79,paquette1986,koester09}, indicating that a process should be at work that counteracts it or replenishes the depleted metals.  
  
In addition to the formation and evolutionary history of these objects, their future outcome is also of theoretical interest. Short-period double WD binaries are candidate progenitors for transient explosive phenomena such as Type~Ia, underluminous .Ia and Ca-rich supernovae \citep{bilsten07,iben84,perets2010,foley2015}, as well as exotic systems such as AM~CVn stars, R~Coronae Borealis (R~CrB), and single subdwarf B/O~stars
 \citep{kilic14am, solheim2010, clayton2013,heber2016}. Moreover, they are expected to be excellent sources of gravitational waves \citep{hermes2012b,kilic2013a} and verification sources for gravitational detectors such as \textit{eLISA} \citep{amaro2012}.

\section{Formation and evolution of ELM~WDs} 
From  a theoretical point of view, an ELM~WD can be formed either through common-envelope evolution or stable Roche-lobe overflow (RLO) mass transfer in a low-mass X-ray binary (LMXB)
or a cataclysmic variable (CV) system. The formation and evolution of low-mass WDs through a stable mass-transfer phase (or by artificially removing envelope mass from its progenitor star)
has been studied intensively over the years \citep[e.g.][]{dribe98,sarna2000,nelson04,althaus2001,panei07,althaus13,itl14,itla14}. 
In comparison, the common-envelope channel is less studied and far more uncertain \citep[e.g.][]{nandez15}. 

Although the majority of ELM~WDs are found in double WD systems \citep{andrews14}, almost all evolutionary calculations that involve stable mass transfer producing an ELM~WD 
consider a neutron star companion (i.e. an LMXB progenitor system). For the structure of the final ELM~WDs, the results of these LMXB calculations can also be applied to CV systems producing ELM~WDs in double WD binaries, as the stellar properties of the produced ELM~WDs do not depend on the mass of their accreting companion, but instead on the initial orbital period and mass of the donor (progenitor) star \citep{nelson04,devito10,itl14}. Only the orbital periods of the produced ELM~WDs will be different.

\subsection{Hydrogen shell flashes and proto-WDs}
After the RLO mass-transfer phase ends, the remaining donor star goes through a so-called (bloated) proto-WD phase in which a significant part of the hydrogen left in the 
envelope is burned through stable hydrogen shell burning. In addition to this, depending on the mass of the proto-WD, its metallicity and the physics included
in the modelling, hydrogen may be burned through short-lived phases of unstable burning through CNO hydrogen shell flashes \citep[e.g.][]{dribe98,althaus2001c,nelson04}.
 
\begin{figure}
	\centering
	\includegraphics[width=\columnwidth]{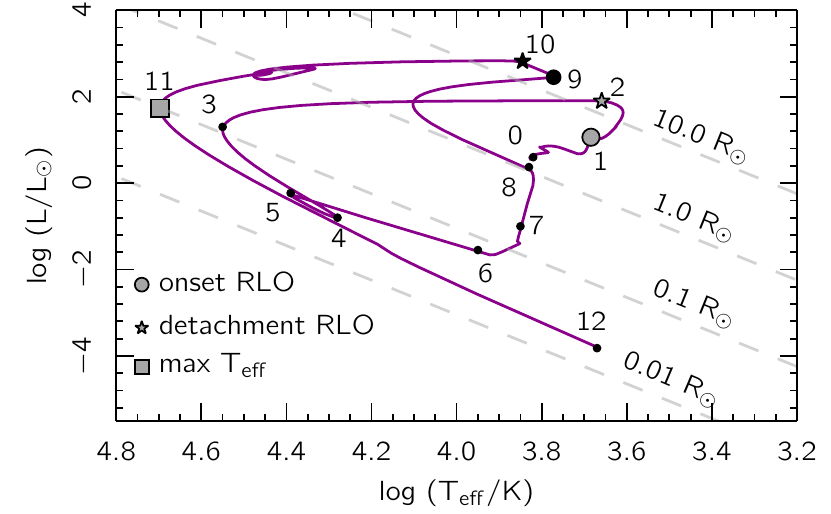}
  	\caption{Hertzsprung-Russel (HR) diagram showing the formation and cooling of a 0.28 M$_{\odot}$ helium WD (produced in an LMXB) that undergoes a hydrogen shell flash. The initial progenitor mass is $1.4\;M_{\odot}$ ($Z=0.02$), the neutron star mass is $1.2\;M_{\odot}$, and the initial orbital period is 5.0~days. See Table~\ref{tab:example_flash} for ages at each stage.}
	\label{fig:example_flash}
\end{figure} 
 
Figure~\ref{fig:example_flash} shows an example of the formation of an ELM~WD through the LMXB channel, including the evolution as a proto-WD as well as its  further cooling. The stellar track is computed from the zero-age main sequence (ZAMS) until the donor star reaches an age of 14 Gyr, points 0 and 12, respectively, in Fig.~\ref{fig:example_flash}. In this case, the star experiences one hydrogen shell flash. 
 After the Roche-lobe detachment (point~2), the proto-WD goes through a phase of contraction 
at almost constant luminosity and increasing effective temperature (between points~2 and 3). The total luminosity is dominated by CNO burning while the contribution due to release of gravitational binding energy from contraction is negligible. When the proto-WD reaches point~3, which is at the beginning of the cooling branch, the temperature in the burning shell is too low to sustain CNO burning, 
therefore the main contribution to the total luminosity is for a while  given by contraction, until the star switches to pp-burning. The unstable burning starts around point~4 and CNO burning becomes dominant again. The increasing energy release during the flash development creates a steep temperature gradient close to the location of maximum energy production. This will give rise to a pulse-driven convection zone within the hydrogen burning shell. After the convection zone is fully developed, the evolution becomes faster (between points~5 and 6). Around point~7, the convection zone reaches the stellar surface, and consequently, its surface chemical composition is altered. The maximum hydrogen luminosity reached during the flash is supplied by the pp-burning, although the onset of the instability is triggered by the CNO cycling. Between points~7 and 8, the lower boundary of the pulse-driven convection zone moves upwards, and at point~8 it completely vanishes. Beyond point~8, the contraction of the inner shells resumes, while the surface layers react by expansion, resulting in a redward motion in the HR-diagram that almost brings the proto-WD back to the red-giant branch. At point~9, the star fills its Roche~lobe again and a short episode of mass transfer is initiated (between points~9 and 10) 
with a high mass-transfer rate that approaches $\sim\!10^{-7}\;M_{\odot}\,{\rm yr}^{-1}$. 
After point~10, the star again evolves towards a high surface temperature at almost constant luminosity, and before reaching the final cooling track, it develops a so-called subflash (near log $T_{\mathrm{eff}}=4.4$). The time intervals for each of the above described phases are shown in Table~\ref{tab:example_flash}.  

\begin{table}
\centering
\caption{Evolution as a function of time for a $0.28\;M_{\odot}$ proto-WD evolving through a hydrogen shell flash, as plotted in Fig.~\ref{fig:example_flash}. The ZAMS is at point~0, and the onset of RLO (the LMXB~phase) is at point~1. The relative age is in comparison to the previous point of evolution and the WD age is with respect to the Roche-lobe detachment (point~2). See text for more details.}
\label{tab:example_flash}
\begin{tabular}{rrr} 
\hline
Point & Relative age (Myr) &  WD age (Myr) \\
\hline		
	0 & --                     & --  \\
	1 & 3330                   & --  \\
	2 &  217                   &  0  \\
    3 & 4.92                   & 4.92 \\
    	4 & 6.63                   &  11.5 \\
    5 & 57.1                   &  68.7 \\
    	6 & 0.0066                 &  68.7 \\
   	7 &  2.4$\times$ 10$^{-5}$ &  68.7 \\
    8 &  3.8$\times$ 10$^{-5}$ &  68.7 \\
    9 &  1.5$\times$ 10$^{-5}$ &  68.7 \\
   10 &  3.1$\times$ 10$^{-5}$ &  68.7 \\
   11 & 0.193                  &  68.9 \\	  
   12 & 10\,700                &  10\,800\\	  
\end{tabular}
\end{table}
 
The proto-WD phase has an associated timescale, $\Delta t_{\rm proto}$, which is the time it takes the star to evolve (and contract) from the Roche-lobe detachment until it reaches its maximum effective temperature on the (final) cooling track. In Fig.~\ref{fig:example_flash} this corresponds to
the time interval during the evolution from point~2 to point~11. The duration of this contraction phase is mainly given by the burning rate of the residual hydrogen in the envelope. For evolved low-mass stars there is a well-known correlation between the degenerate core mass and its luminosity \citep{refsdal71}. Therefore, after Roche-lobe detachment, the rate at which the residual  hydrogen in the envelope is consumed is directly proportional to the luminosity and thus increases strongly with $M_{\mathrm{WD}}$. A detailed analysis of the  dependence of $\Delta t_{\rm proto}$ on the mass of the WD is given in \citet{itla14}. 
This timescale is especially important in MSP systems and should be added to the optically determined cooling age of the WD to yield the true age of the recycled radio pulsar.
Unfortunately, the true age of a recycled pulsar cannot be determined from its characteristic spin-down age as this method has proved unreliable by a factor of 10 or more \citep{camilo94,lorimer95,tauris12,tlk12}.

Hydrogen shell flashes occur in a range of proto-WD masses that is dependent on the metallicity and whether or not element diffusion is included in the modelling. 
The lower mass limit for flashes, $M_{\mathrm{flash, min}}$, is determined by the size of the burning shell, such that if $M_{\mathrm{proto-WD}}$ < M$_{\mathrm{flash, min}}$, 
then the shell is too thick to trigger unstable hydrogen burning.
The upper mass limit, $M_{\mathrm{flash, max}}$, is determined by the cooling time of the burning shell, which needs to be long enough to avoid an extinction 
of the shell before the instability is fully established \citep[if $M_{\mathrm{proto-WD}}$ > $M_{\mathrm{flash, max}}$, this condition is not fulfilled, cf.][]{dribe98}. 
These conditions are altered when element diffusion is included \citep{althaus2001c}, and this issue is investigated more carefully in Sect.~\ref{results}.  
    
\subsection{Age dichotomy in helium WD cooling?}\label{subsec:intro_dichotomy}
The occurrence of hydrogen shell flashes, when element diffusion is taken into account, has been found to be responsible for a dichotomy in the cooling ages of helium WDs \citep{althaus2001,kerkwijk05,althaus13,Bassa16}.
The occurrence of flashes in relatively massive helium WDs ($>0.2\;M_{\odot}$), with initially thin hydrogen envelopes, leaves behind an even  thinner envelope, giving rise to
relatively fast cooling. On the other hand, less massive (proto) helium WDs ($<0.2\;M_{\odot}$) have thicker hydrogen envelopes after RLO, resulting in stable shell hydrogen burning,  
and will therefore continue residual hydrogen burning on the cooling track on a long timescale. 

Recently, \citet{itla14} found no evidence for such a dichotomy in the case of thermal evolution of proto-WDs but rather a smooth transition with the mass of the WD.      
The authors showed that the thermal evolution timescale mainly depends on the proto-helium WD luminosity, which in turn depends on the mass of the proto-WD 
and not on the occurrence of hydrogen shell flashes. These new findings questioned whether  a dichotomy exists in the cooling ages of ELM~WDs and if the responsible process might be the occurrence of hydrogen shell flashes. 

\subsection{Aims of this investigation of ELM~WDs}
The focus of this paper is on the proto-WD phase of ELM~WDs, which are investigated through a series of binary stellar evolution calculations of LMXBs. 
The following aspects are addressed: 
(i) the hydrogen envelope mass as a result of binary evolution, 
(ii) the role played by rotational mixing in the evolution of ELM~proto-WDs, 
(iii) the influence of element diffusion and rotation on $\Delta t_{\rm proto}$ as well as on the cooling timescale, 
(iv)  the existence of a dichotomy in the cooling ages of ELM~WDs as a result of the occurrence of hydrogen shell flashes, 
(v) the presence of metals in the atmospheres of proto-WDs, and 
(vi) the relation between the mass of a proto-WD and its orbital period at the end of the mass-transfer phase.  
All these aspects are addressed not only as a function of the proto-WD mass, but also as a function of metallicity. Answering these open questions is essential for understanding the formation of ELM~WDs and their age determination, for providing accurate models for astroseismology calculations, and for determining the correct age of MSP binaries. This work extends the previous work by \citet{itla14} by including element diffusion and rotational mixing in the evolution of the donor star and during the proto-WD and the WD cooling phase. Moreover, the study is extended  to include the effect of metallicity as well, for which we investigate
four metallicities: $Z=0.02$, $0.01$, $0.001$, and $0.0002$.  

%%%%%%%%%%%%%%%%%%%%%%%%%%%%%%%%%%%%%%%%%%%%%%%%%%%%%%%%%%%%%%%%%%%%%%%%%%%%%%%%%%%%%%%%%%%%%%%%%%%%%%%%%%%%
\section{Numerical methods}\label{numerical_methods}
The evolutionary tracks presented in this paper are calculated using the publicly available binary stellar evolution code MESA, 
version 7624 \citep{mesa10,mesa13,mesa3}. The nuclear network used is \texttt{cno$\_$extras.net} and accounts for the CNO burning with the following isotopes: 
$^{1}$H, $^{3}$He, $^4$He, $^{12}$C, $^{13}$C, $^{13}$N, $^{14}$N, $^{15}$N, $^{14}$O, $^{15}$O, $^{16}$O, $^{17}$O, $^{18}$O,  $^{17}$F, $^{18}$F, $^{19}$F,$^{18}$Ne, $^{19}$Ne, $^{20}$Ne, $^{22}$Mg and $^{24}$Mg.  
Radiative opacities are taken from \citet{ferguson05} for $2.7\leq \log T \leq4.5$ and OPAL \citep{iglesias93,iglesias96} for $3.75\leq \log T \leq8.7$
and conductive opacities are adopted from \citet{cassisi07}.
Convective regions are treated using the mixing-length theory (MLT) in the \citet{henyey65} formulation with $\alpha_{\rm {MLT}}=2.0$. 
Transport of angular momentum is treated as a diffusive process which results in rigid rotation in convective zones. The boundaries of convective regions are determined using the Schwarzschild criterion. A step function overshooting extends the mixing region 
for 0.2~pressure scale heights beyond the convective boundary during core H-burning.

 We here refer to element diffusion as the physical mechanism for mixing of chemical elements that is due to pressure gradients (or gravity, i.e. gravitational settling), temperature
(thermal diffusion) and composition gradients (chemical diffusion). Gravitational settling tends to concentrate heavier elements towards  the centre of the star. Thermal diffusion generally acts in the same direction, although to a lesser degree, by bringing highly charged and more massive species  towards the hottest region of the star (its centre). Chemical diffusion, on the other hand, has the opposite effect \citep[e.g.][]{iben85, thoul94}. MESA includes the treatment of element diffusion through gravitational settling, chemical and thermal diffusion \citep{thoul94}, and radiative accelerations \citep{hu11}. Radiative forces are proportional to the reciprocal of the temperature and are thus negligible in hot regions where nuclear burning is of importance. In addition, calculating
these forces is computationally demanding. We therefore here neglected the effects of radiative levitation (which is important for determining photospheric composition
of hot WDs \citep{fontaine79}. The detailed description of how element diffusion is implemented in MESA 
can be found in \citet{mesa3}. We take into account the effects of element diffusion due to gravitational settling and chemical and thermal diffusion 
for the following elements  $^1$H, $^{3}$He, $^{4}$He, $^{12}$C, $^{13}$C, $^{14}$N, $^{16}$O, $^{20}$N, $^{24}$Mg, and $^{40}$Ca.\\
MESA includes the effects of the centrifugal force on stellar structure, chemical mixing, and transport of angular momentum that is  due to rotationally induced 
hydrodynamic  and secular instabilities as described in \cite{heger00}. Here, we take into account the mixing due to dynamical shear instability, secular shear instability, 
Eddington-sweet circulation, and Goldreich-Schubert-Fricke instability with a mixing efficiency factor of $f_{c}=1/30$ \citep{heger00}. The mixing of angular momentum that is due to dynamo-generated magnetic fields in radiative zones is also included \citep{spruit2002,heger05} as is the angular momentum transport due to electron viscosity \citep{itoh87}.
A decrease of the mean molecular weight with radius has a damping effect on mixing processes driven by rotation or even prevents these   from occurring. The strength of this effect is regulated by the parameter f$_{\mu}$, for which we follow \citet{heger00} and set  f$_{\mu}$=0.05.

The initial metallicity was set to $Z=0.02$ (Y=0.28), with initial abundances from \citet{grevesse98}. The lower metallicities were  obtained by scaling both X and Y by the same factor such that $X+Y+Z=1$.  For the WD evolution and for $T_{\rm{eff}}<10\;000\;{\rm K}$, the outer boundary conditions were derived using non-grey model atmospheres \citep{rohrmann12}.

To calculate the rate of change of orbital angular momentum, we took  into account contributions from gravitational wave radiation, mass loss, magnetic braking, and spin orbit couplings: 
\begin{equation}  
    \dot{J}_{\rm orb} = \dot{J}_{\rm gwr} + \dot{J}_{\rm ml} +\dot{J}_{ \rm mb} + \dot{J}_{\rm ls}\,,
\label{orbital_angular_momentum}
\end{equation}
as described in \citet{mesa3}.
The contribution of spin-orbit couplings to $\dot{J}_{\mathrm{orb}}$ was computed by demanding conservation of total angular momentum (except for losses due to gravitational wave radiation, magnetic braking, and mass loss), that is, changes in spin angular momentum were compensated for by changing the orbital angular momentum. The initial rotation velocity of the donor star was set by requiring that its spin period be synchronized with the initial orbital period. The time evolution of the angular velocity of the donor star is given by  
\begin{equation}
  \frac{d\Omega_{i}}{dt}=\frac{\Omega_{\rm{orb}}-\Omega_{i}}{\tau_{\rm{sync}}},
\end{equation}
where $\Omega_{i}$ is the angular velocity of cell $i$ \citep{detmes2008}. The synchronization time, $\tau_{\rm sync}$ was calculated using the formalism of tidal effects from \cite{hurley2002} 
and depends on whether the envelope is convective or radiative.  

\subsection{Grid of models}
To produce our grid of models, we followed the detailed binary evolution of the donor star from the ZAMS until it reached an age of 14~Gyr. The neutron star was treated as a point mass. The final outcome of these LMXB systems is very sensitive to the initial orbital period and to the treatment of orbital angular momentum loss \citep[e.g.][and references therein]{itl14}. 

We calculated binary tracks for four metallicities: $Z=0.02$, 0.01, 0.001, and 0.0002. For each metallicity, the models were divided into three categories: 
(i) basic models (with no diffusion nor rotation), 
(ii) diffusion models (with element diffusion) and,  
(iii) diffusion+rotation models (with element diffusion plus rotation). 
In both the diffusion and diffusion+rotation models, we included the effects of centrifugal forces and angular momentum transport, which means that these two models only differ by the presence of rotational mixing in the rotation models.

For $Z=0.02$, the initial binary configuration has a $1.4\;M_{\odot}$ donor star and a $1.2\;M_{\odot}$ neutron star accretor. For all the other metallicities, our models were calculated with a 
$1.0\;M_{\odot}$ donor star and a $1.4\;M_{\odot}$ neutron star (to facilitate direct comparison with previous work in the literature, see Sect.~\ref{subsec:compare}). 
All the models were computed using a magnetic braking index of $\gamma=4$, and we assumed that 30~per~cent of the transferred mass is  ejected from the neutron star as a fast wind carrying its specific orbital angular momentum. We note that the structure of the ELM~WDs is not sensitive to the above choices of mass-transfer parameters which only affect their final orbital periods.
A comprehensive study of the influence of the magnetic braking index and the accretion efficiency on LMXB evolution can be found in \citet{itl14}. We point out again that we here refer to the mass of the proto-WD as being the (bloated) donor star mass at the end of the RLO mass-transfer phase (before the occurrence of flashes, which can lead to additional mass-transfer episodes), and the mass of the WD as being the mass at the beginning of the cooling track.  
We calculated models just above the bifurcation period, which is defined as the shortest initial orbital period that produces a WD \citep[e.g.][and references therein]{itl14}.

%%%%%%%%%%%%%%%%%%%%%%%%%%%%%%%%%%%%%%%%%%%%%%%%%%%%%%%%%%%%%%%%%%%%%%%%%%%%%%%%%%%%%%%%%%%%%%%%%%%%%%%%%%%%
\section{Results}\label{results}
\subsection{General effects of element diffusion and rotational mixing}\label{sec:element_diffusion_general}
In the context of low-mass helium WDs, element diffusion was investigated in detail over the past few years by the  La~Plata group \citep[e.g.][]{althaus2000,serenelli01,althaus2001,althaus2001c,althaus2001b,panei07,althaus2009,althaus13} using the stellar evolution code LPCODE for various ranges of helium WD masses and metallicity. To the best of our knowledge, there is only one other study that used MESA for low-mass helium WDs \citep{gautschy13}. Our models include element diffusion from the ZAMS and not only from the proto-WD phase, as in the previous works. Moreover, for the first time, we investigate in detail the role played by rotational mixing in addition to element diffusion in the evolution of ELM~WDs.  
\begin{figure}
\centering
\includegraphics[width=\columnwidth]{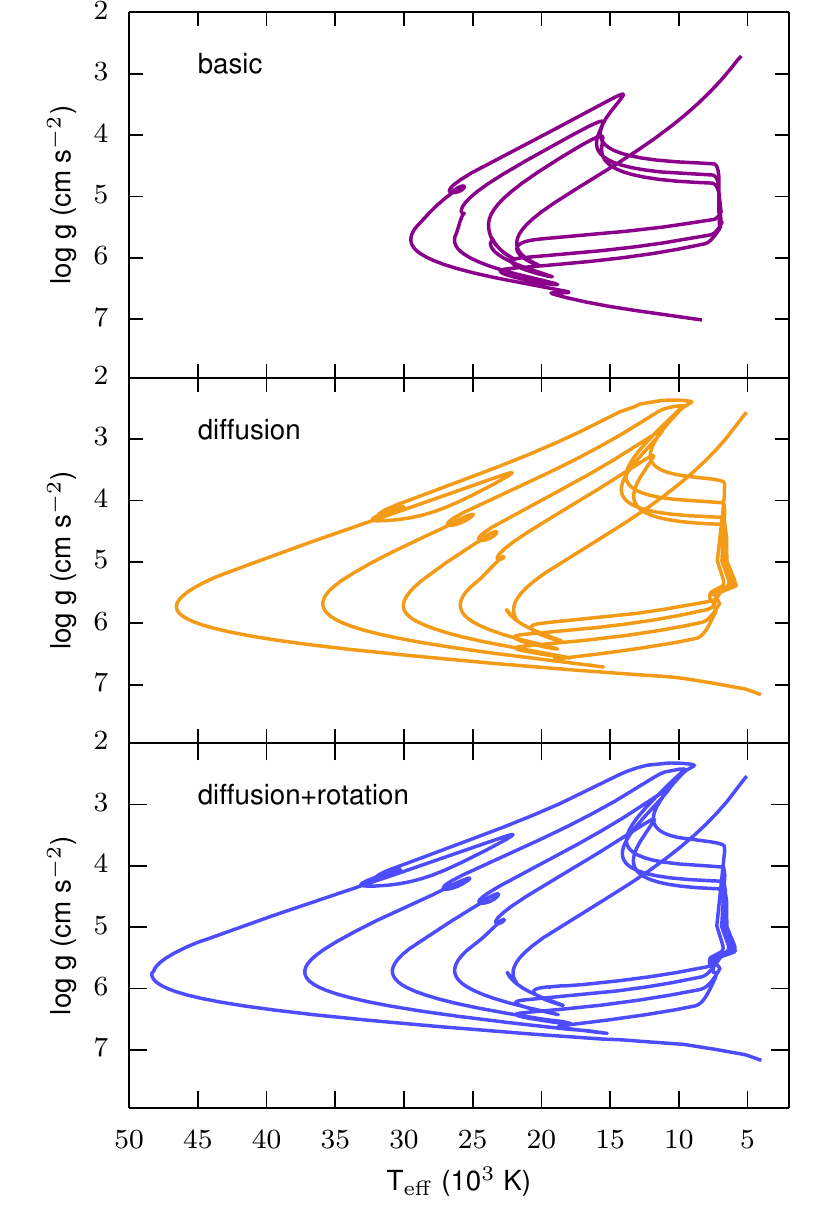}
\caption{Post-RLO evolution of surface gravity versus effective temperature for a $\sim\!0.23\;M_{\odot}$ proto-WD produced from an LMXB donor star with $Z=0.01$.
Three different models are shown: basic configuration (top panel), diffusion configuration (middle panel) and diffusion+rotation configuration (bottom panel). 
Note that the computations are stopped when the age of the model star reaches 14~Gyr (since ZAMS).}
\label{fig:compare_logg_teff}
\end{figure} 

Element diffusion has a strong effect on the surface composition of a proto-WD and on the chemical profile deep inside the star close to the helium core. 
At the surface,  gravitational settling increases the hydrogen abundance given that hydrogen is the lightest element. Close to the helium core boundary, chemical diffusion tends to smooth it out by mixing the hydrogen  downwards into hotter layers because a large hydrogen abundance gradient exists. 
It has been shown that this hydrogen tail promotes the occurrence of hydrogen shell flashes \citep[e.g.][]{althaus2001}. Moreover, when element diffusion is included, a  proto-WD experiences more flashes  than when diffusion  is neglected \citep[e.g.][]{althaus2001}.
 The WD mass interval in which they occur is also changed compared to the case when element diffusion is ignored.  The number of flashes and other information for all the models studied in this work are given in Appendix~\ref{appendix:tables}. 

\begin{figure}
\centering
\includegraphics[width=\columnwidth]{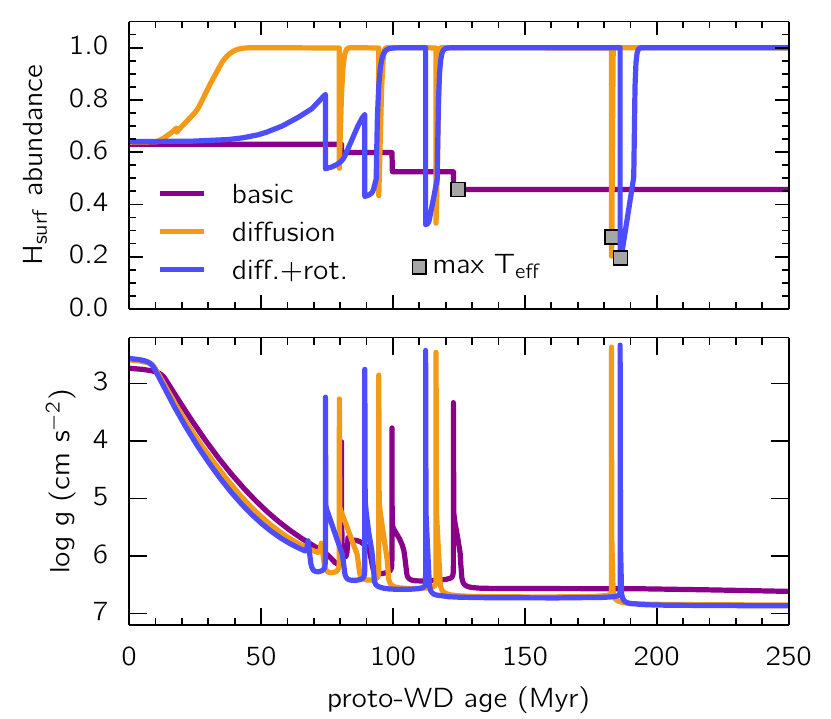}
\caption{Evolution of hydrogen surface abundance (top panel) and log $g$ (bottom panel) for the three proto-WDs shown in Fig.~\ref{fig:compare_logg_teff} illustrating the effect of gravitational settling, rotational mixing and the mixing due to convection zones developed during the hydrogen shell flashes  on the surface composition of these objects. }
\label{fig:hydrogen_surface}
\end{figure}

Figure~\ref{fig:compare_logg_teff} shows the evolution of surface gravity versus effective temperature for a proto-WD of $\sim\!0.23\;M_{\odot}$ obtained from the following three model configurations: basic, diffusion, and diffusion+rotation. The basic model  experiences three hydrogen shell flashes, while the diffusion and the diffusion+rotation models experience one additional flash. The radial expansion following the CNO burning is more pronounced when  element diffusion is included, in some cases  leading to additional episodes of RLO. In general, the models with diffusion and diffusion+rotation behave in a very similar way. 
\begin{figure*}
\centering
\subfloat[]{\includegraphics[width=0.48\textwidth]{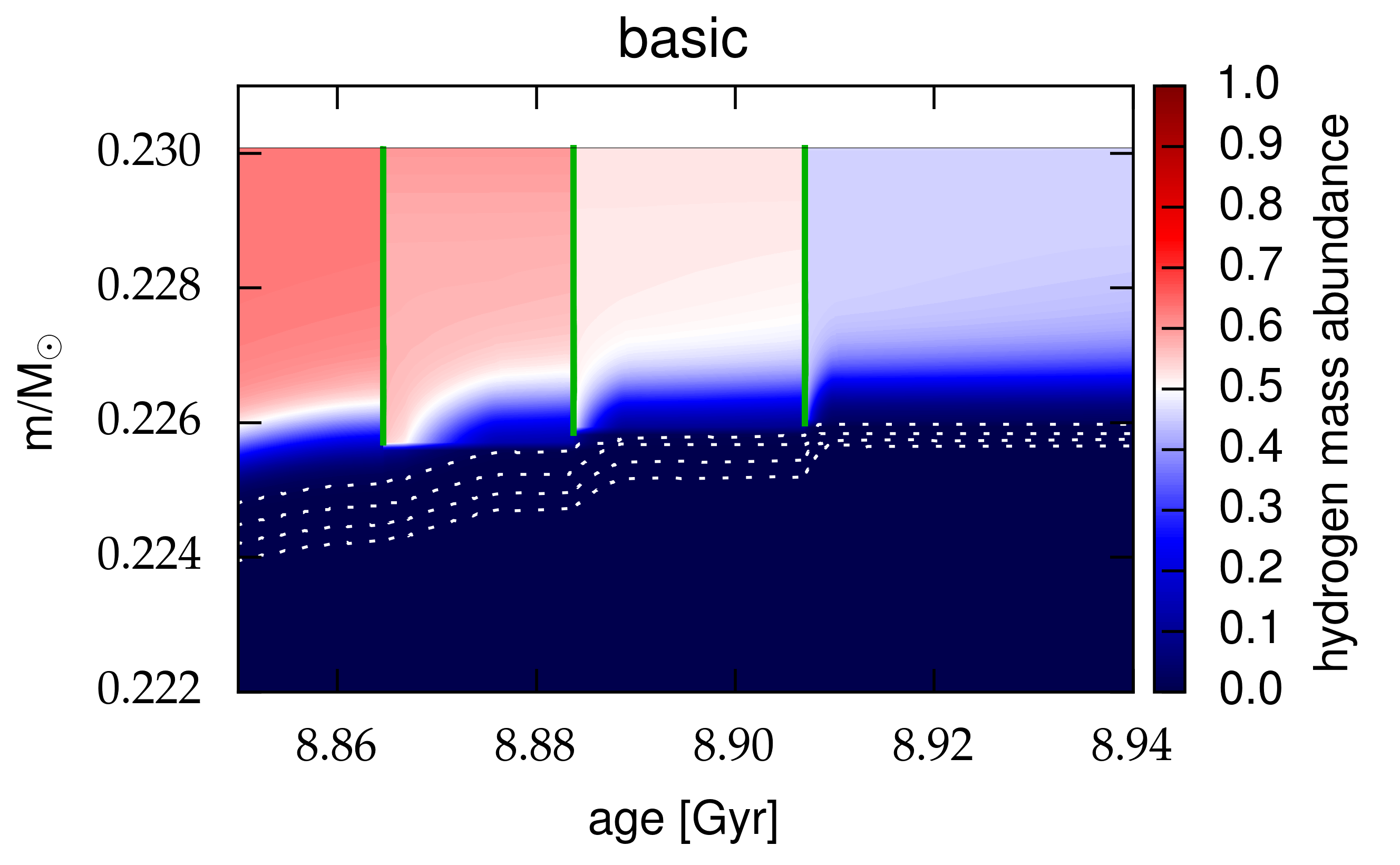}}\\
\subfloat[]{\includegraphics[width=0.48\textwidth]{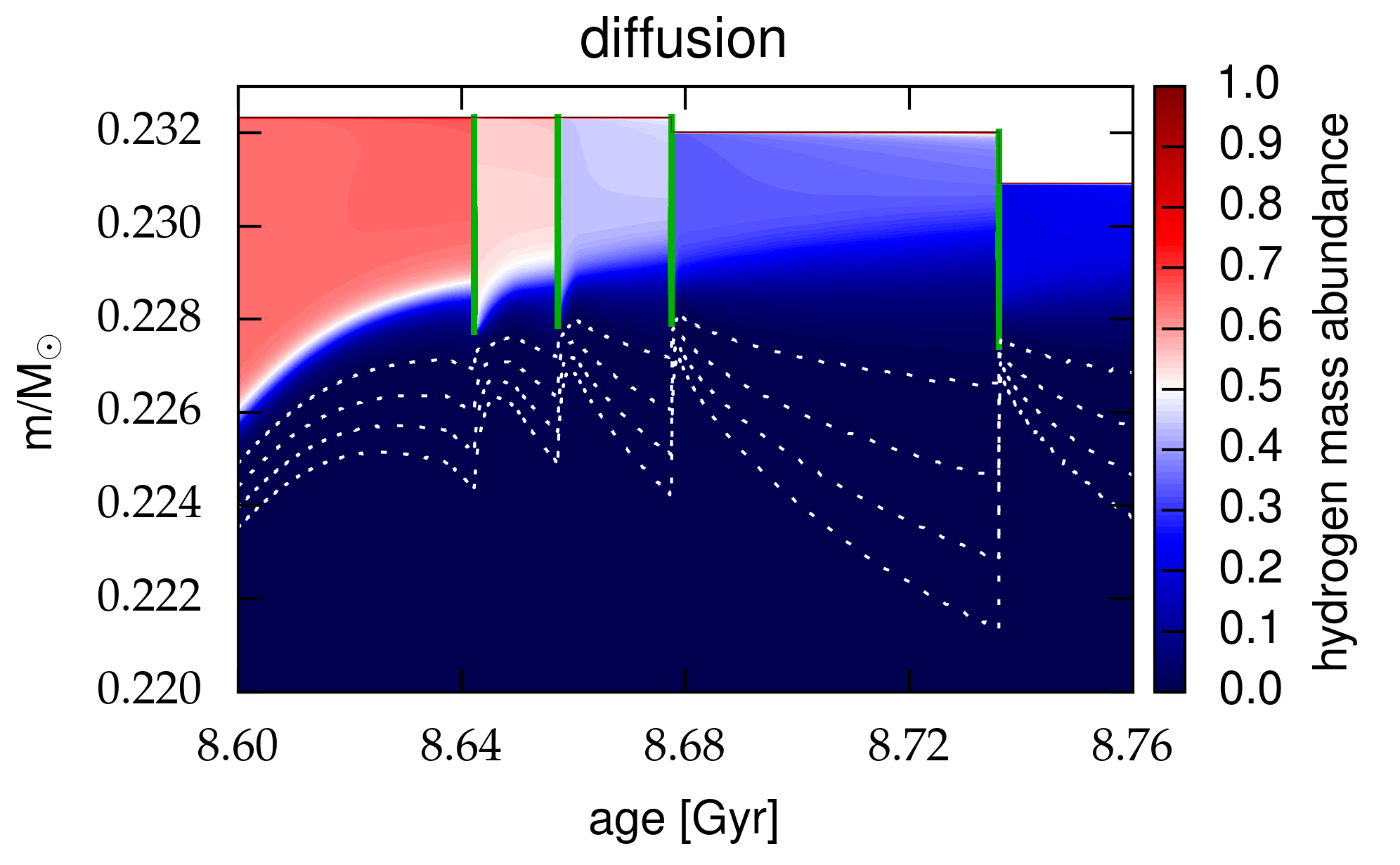}}\quad
\subfloat[]{\includegraphics[width=0.48\textwidth]{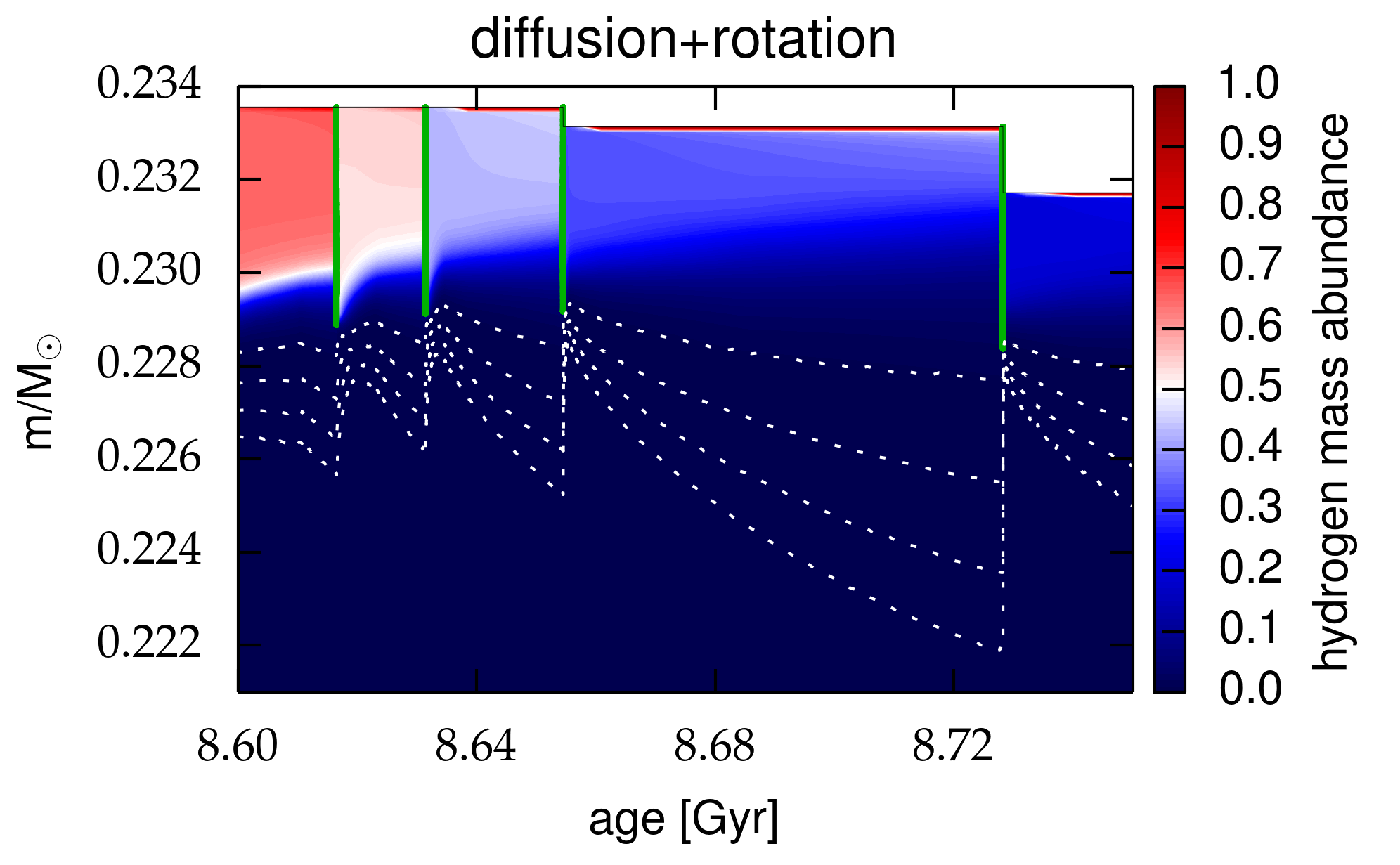}}\\
\caption{Kippenhahn diagrams showing the proto-WD phase for the same systems as in 
Figure~\ref{fig:compare_logg_teff}. The plots show cross sections of the outer $\sim$ $0.01\;M_{\odot}$ envelope of the proto-WD in mass coordinates, along the $y$-axis, as a function of stellar age on the $x$-axis (relative to the ZAMS age). The green  areas denote zones with convection; the  dotted white lines define lines of constant hydrogen abundance, 10$^{-2}$ to 10$^{-5}$ (from top to bottom). 
The intensity of the blue and red colour indicates the hydrogen abundance by mass fraction, as shown on the colour scale to the right. As a result of different input physics, the proto-WDs have slightly different masses and ages. See text for details. }
\label{fig:kipp_plots}
\end{figure*}

\begin{figure}
\centering
\includegraphics[width=\columnwidth]{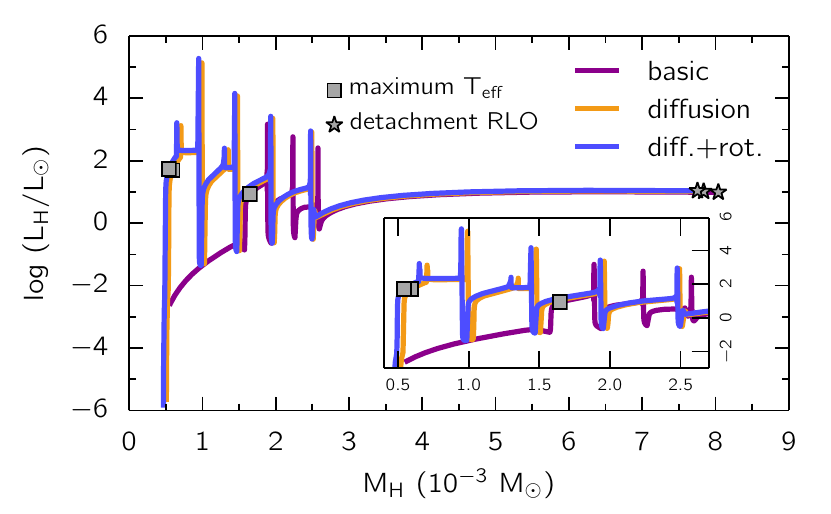}
\caption{Hydrogen burning luminosity versus hydrogen envelope mass, for the same systems as in Fig.~\ref{fig:compare_logg_teff}. The grey stars represent the moment of Roche-lobe detachment 
while the grey squares denote the point of maximum effective surface temperature. The evolution is from the right to the left.}
\label{fig:luminosity_hydrogen}
\end{figure}

Figure~\ref{fig:hydrogen_surface}  shows  the evolution of hydrogen surface abundance (top panel) and log~$g$ (bottom panel) for the same (proto)WDs as in Fig.~\ref{fig:compare_logg_teff}. As already mentioned, gravitational settling changes the surface abundances. All the elements 
heavier than hydrogen sink below the surface, leaving a pure hydrogen envelope behind. 
When rotational mixing is included, gravitational settling and rotational mixing compete with each other to determine the chemical composition of the surface. At the beginning of the proto-WD phase, rotational mixing dominates. However, the surface gravity of the proto-WD increases with time, while the efficiency of rotational mixing decreases, as described in Sect.~\ref{rotational_mixing}. Thus, in  later phases of the evolution, the gravitational settling overcomes the mixing induced by rotation. By the beginning of the last flash, the surface structure of the model that only includes element diffusion is nearly identical to the structure of the model  that  includes both element diffusion and rotational mixing. 
Helium in the envelopes of ELM~proto-WDs is a crucial ingredient for exciting pulsation modes through the $\kappa-$mechanism, as shown by \citet{jeffsaio2013} and \citet{corsico16} for radial and nonradial modes. \cite{gianninas2016} recently provided the first empirical evidence that pulsations in ELM proto-WDs can only occur when a significant amount of helium is present in their atmospheres. In contrast with evolutionary models  that only include element diffusion, our new evolutionary models including rotational mixing produce proto-WDs that have mixed He/H envelopes during most of their evolution before settling on the cooling track.   \\
Another effect of element diffusion, resulting  mainly from the competition between chemical and thermal diffusion, is the development of a hydrogen tail that reaches down into the hot helium-rich layers,    as shown in Fig. \ref{fig:kipp_plots}. This effect is responsible for the larger number of flashes compared to the case where element diffusion is ignored (basic model).  
Rotational mixing is seen not to change the chemical structure of the deep layers. 
This can also be concluded from the very similar behaviour in terms of the number of flashes and the structure of the flashes in the case that includes both diffusion and rotation compared to the case that only includes element diffusion, cf. Figs.~\ref{fig:compare_logg_teff} and \ref{fig:luminosity_hydrogen}.

In Fig.~\ref{fig:luminosity_hydrogen} we plot the luminosity produced by hydrogen burning versus the hydrogen envelope mass. For all three models, around 70~per~cent of the  hydrogen remaining from the end of the LMXB~phase (Roche-lobe detachment) is processed before the occurrence of flashes while the bloated proto-WD crosses the HR--diagram. The occurrence of additional flashes, which applies to the cases where element diffusion is included, reduces the hydrogen envelope mass available on the cooling track (i.e. after reaching maximum $T_{\rm eff}$, marked by squares in Fig.~\ref{fig:luminosity_hydrogen}) by a factor of $\sim\!3$ compared with the basic model. 
The basic model still experiences significant residual hydrogen burning on the cooling track.   
As a result, the basic model only cools down to a temperature of $T_{\rm eff}\approx 8400\;K$  within 14~Gyr (since the ZAMS), while the two models that include diffusion will cool down to roughly $T_{\mathrm{eff}}\approx 4000\;K$ (cf. Fig.~\ref{fig:compare_logg_teff}). 
The cooling properties of the ELM~WDs are discussed in more detail in Sect.~\ref{cooling_track}.
\begin{figure}
\centering
\includegraphics[width=\columnwidth]{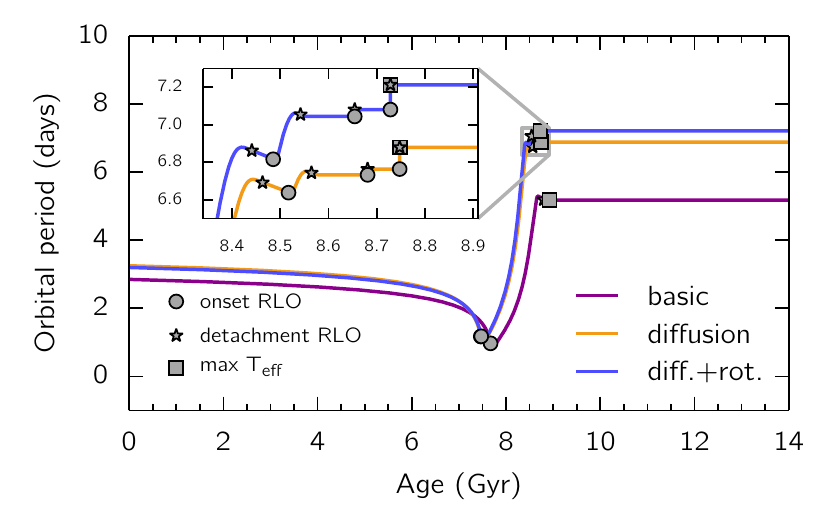}
\caption{Orbital period evolution for the same models as in Fig.~\ref{fig:compare_logg_teff}. The grey circles represent the onset of the mass transfer, the grey stars represent Roche-lobe detachment and the grey squares mark the maximum $T_{\mathrm{eff}}$. }
\label{fig:compare_orbital_period}
\end{figure}

\begin{figure*}
\centering
\includegraphics[width=\textwidth]{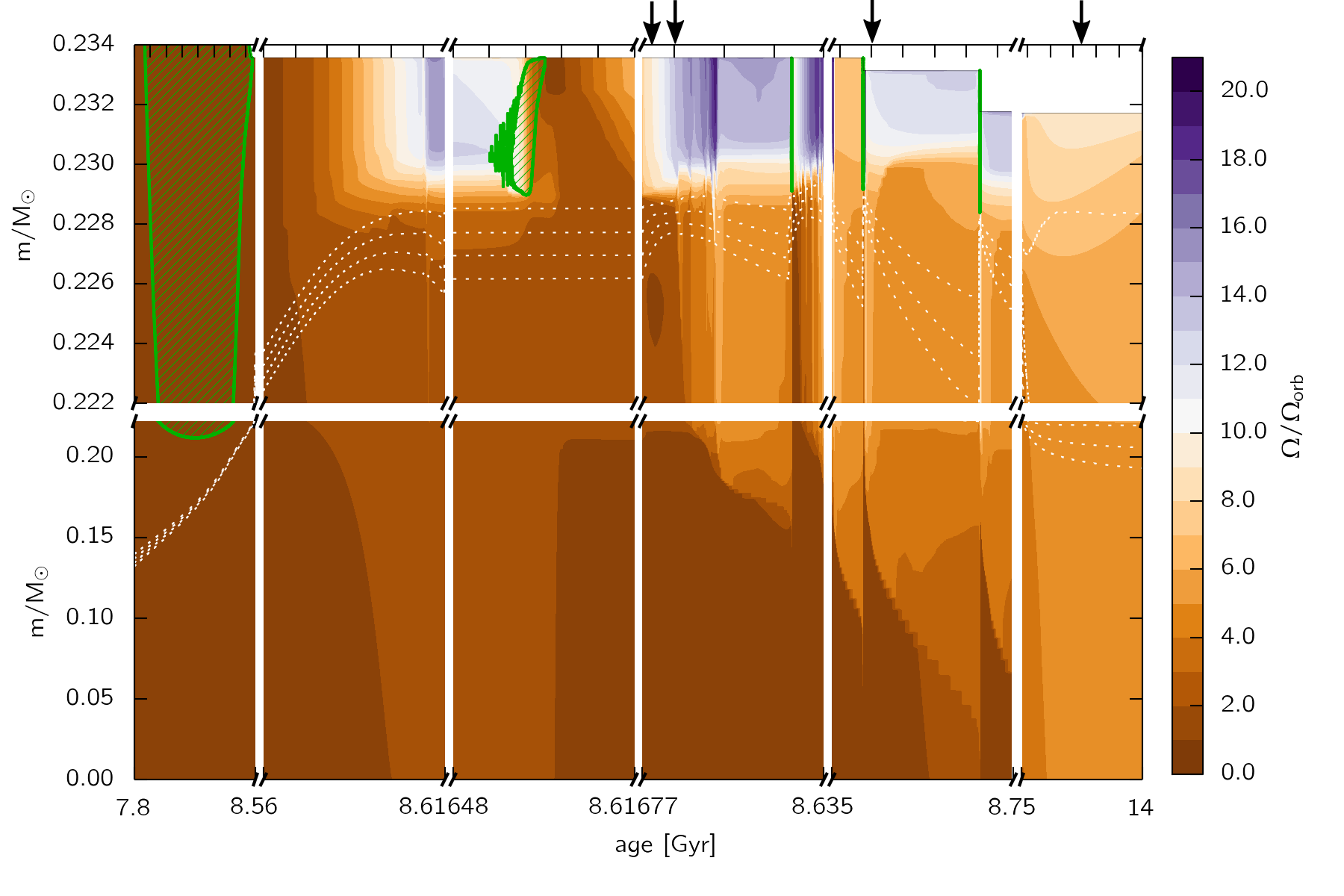}
\caption{Kippenhahn diagram for the same proto-WD as in Fig.~\ref{fig:kipp_plots}, including both element diffusion and rotational mixing. The intensity of the orange and indigo colour indicates the ratio of the spin angular velocity to the orbital angular velocity,   $\Omega/\Omega_{\mathrm{orb}}$, as shown on the colour scale to the right. The green areas and the dotted  white lines have the same meaning as in Fig.~\ref{fig:kipp_plots}. The black arrows point to the position of the profiles in Fig.~\ref{fig:rotation_mechanisms}.}  
\label{fig:omega_paper}
\end{figure*}

The orbital evolution of the models described above is shown in Fig.~\ref{fig:compare_orbital_period}. One difference between the three models is that those with element diffusion (and rotation) require a longer initial orbital period to form approximately the same proto-WD.  
\begin{figure}[ht!]
\centering
\includegraphics[width=\columnwidth]{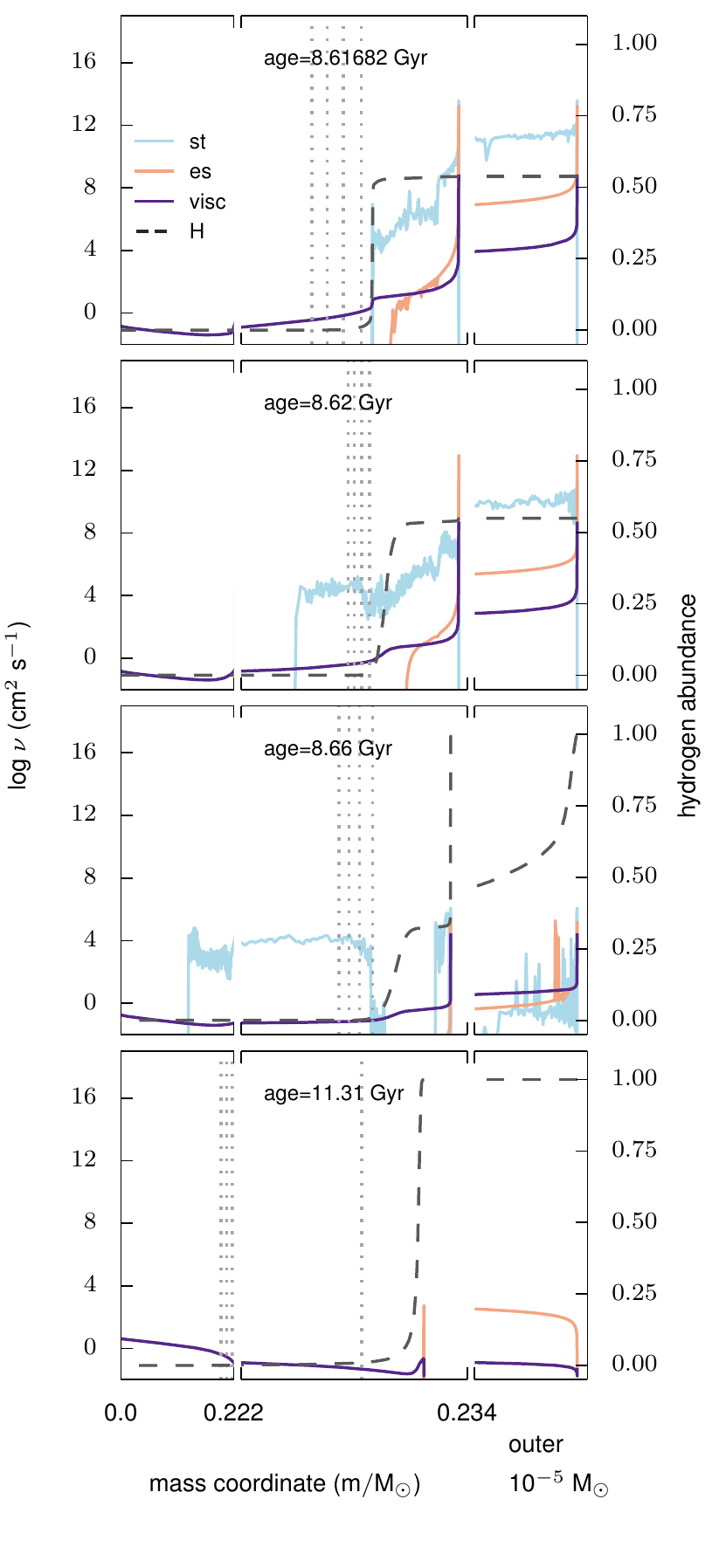}
\caption{ Angular momentum diffusion coefficient, $\nu$ as function of mass coordinate from the centre (left) to the surface of the star (right) for the same proto-WD as in Fig.~\ref{fig:omega_paper}.  Processes included are Spruit-Tayler dynamo (st),  Eddington-Sweet circulation (es), and electron viscosity (visc). Each panel (from top to bottom) corresponds to a profile as marked by the black arrows in Fig. \ref{fig:omega_paper} (from left to right). The grey dotted lines define lines of constant hydrogen abundance, from 10$^{-5}$ to 10$^{-2}$, while the black dashed line represents the hydrogen mass abundance.}
\label{fig:rotation_mechanisms}
\end{figure}
The orbital period at the Roche-lobe detachment is $\sim\!7.05\;{\rm days}$ for the model  that includes diffusion and rotation, $\sim\!6.73\;{\rm days}$ for the model with diffusion only, and $\sim\!5.19\;{\rm days}$ for the basic model. As the diffusion-induced flashes are stronger and because almost every flash causes the star to expand and fill its Roche lobe again, the mass-transfer episodes widen the orbit during each flash. In the end, this effect accounts for an increase of   a few per~cent in the orbital period. 
\subsubsection{Rotational mixing}\label{rotational_mixing}
As shown in Fig.~\ref{fig:omega_paper}, the tidal coupling is strong enough to completely synchronize the donor with the orbit  up until the end of the LMXB~phase. This changes dramatically after detachment from the Roche lobe; while the helium core barely contracts and spins up only slightly above the orbital frequency, the extended hydrogen envelope spins up significantly during contraction, resulting in strong shear at the core-envelope boundary. As hydrogen flashes develop and the star expands and then contracts, the envelope successively spins down and up, with the rotational period at the surface of the proto-WD at its maximum being up to 20 times shorter  than the orbital period. Even though the proto-WD expands back and fills its Roche lobe during flashes, these phases are very short. The convective layers developed during the flashes disappear well before the next phase of Roche--lobe overflow such that tidal synchronization past the LMXB~phase is negligible. 

Although a strong shear is developed, the composition gradients that help stabilize the instabilities driven by rotation prevent the mixing of elements and angular momentum from the envelope to the core. This is shown in Fig.~\ref{fig:rotation_mechanisms}, where the different processes contributing to the angular momentum diffusion coefficient, $\nu$ are shown at four different times. As depicted in the first panel, there is a very steep H-gradient immediately after a flash (and also after detachment from the LMXB phase)  that completely prevents mixing to the core. As burning proceeds between flashes (second and third panels in Fig.~\ref{fig:rotation_mechanisms}), the H-gradient is softened and starts to move outwards in mass, which allows some angular momentum to be transported to the core mainly through magnetic torques from the Spruit-Tayler dynamo. Finally, after settling on the cooling track (final panels panel in Figs.~\ref{fig:omega_paper} and \ref{fig:rotation_mechanisms}), most of the remaining hydrogen has been burnt, and angular momentum has mixed efficiently between the envelope and the core. Despite the small mass of the envelope relative to the total WD mass, this results in the  WD  having a  spin period  more than  four times shorter (i.e. faster)  than its orbital period.\\
Because we have assumed that magnetic torques do not contribute to the mixing of elements, rotational mixing in our models barely affects the formation of the hydrogen tail due to element diffusion, and thus has a weak effect on the strength and the occurrence of flashes. At the surface, however, the fast rotation induces element mixing through  Eddington-Sweet circulation (see Fig.~\ref{fig:rotation_mechanisms}), which counteracts the rapid settling of elements heavier than hydrogen.
\begin{figure}
\centering
\includegraphics[width=\columnwidth]{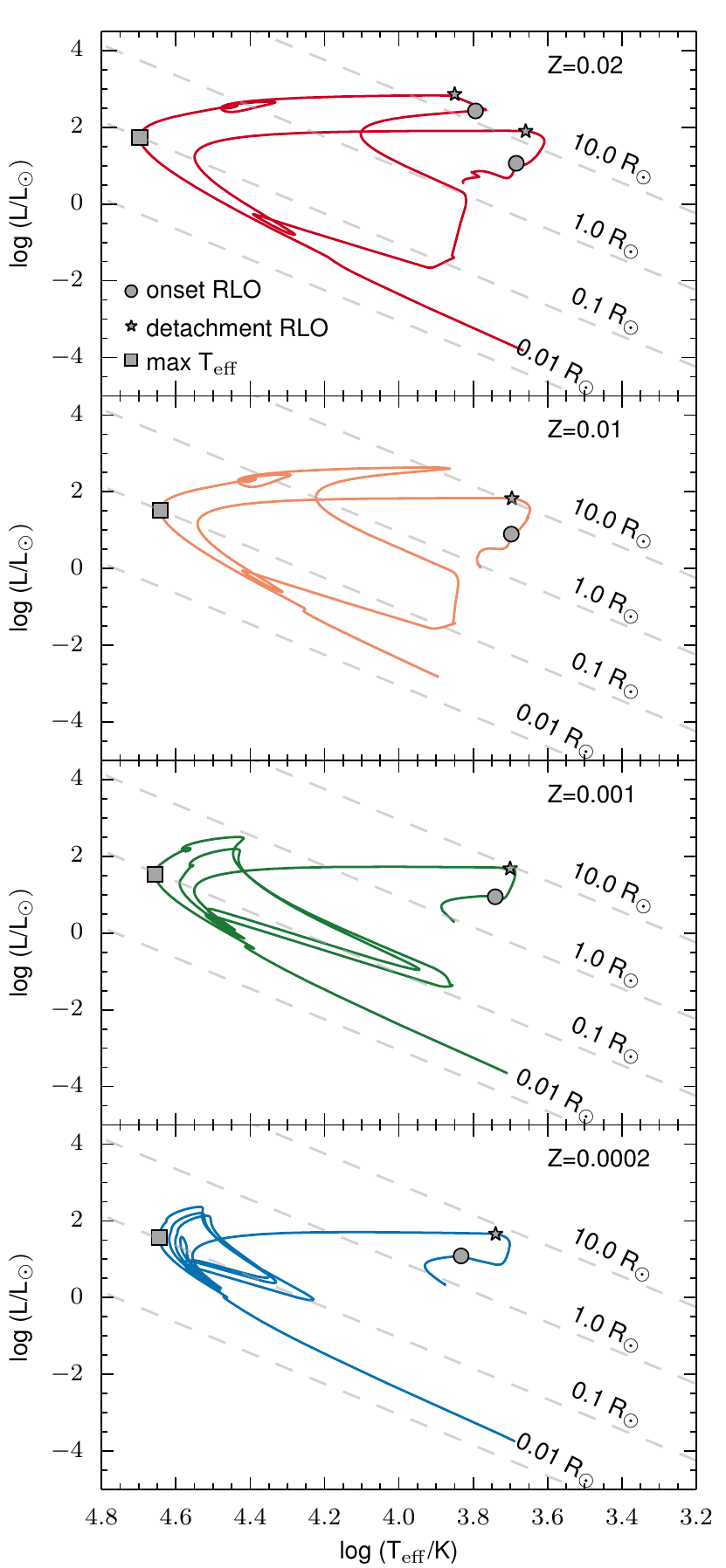}
\caption{HR-diagram showing the formation and evolution of a $\sim\!0.28\;M_{\odot}$ WD from progenitors with different metallicities. The grey symbols represent the beginning of  RLO (circles), the end of RLO (stars) and the maximum $T_{\mathrm{eff}}$ (squares). The grey dashed lines represent lines of constant radius.}
\label{fig:hr_metallicity}
\end{figure}

\subsection{Effect of metallicity }\label{metallicity_effect}
For a given stellar mass, decreasing the metallicity produces a decrease in the radiative opacity which has an impact on the stellar evolution. The ZAMS and the RGB-phase are shifted towards the blue region of the HR-diagram, with the luminosity and effective temperature being 
higher during these phases than for models at solar metallicity. In other words, at low metallicity stars tend to be hotter, have smaller radii, and evolve more quickly than their high-metallicity counterparts. An immediate consequence of lower metallicities is therefore a higher mass of the helium WD formed through the LMXB~phase at a given initial orbital period because the Roche lobe is filled  at a more advanced stage in the evolution as a result  of the smaller radius.   

As previously demonstrated by \citet{sarb2002} and \citet{nelson04}, the threshold mass for the occurrence of hydrogen shell flashes increases with lower metallicity. This is confirmed by our calculations, as shown in Table~\ref{tab:flashes}. For example, for the basic models (without diffusion and rotation), the minimum mass above which the flashes occur is $\sim\!0.21\;M_{\odot}$ for $Z=0.02$, 
$\sim\!0.22\;M_{\odot}$ for $Z=0.01$, $\sim\!0.25\;M_{\odot}$ for $Z=0.001$, and $\sim\!0.28\;M_{\odot}$ for $Z=0.0002$. 

Models with element diffusion have lower threshold values for flashes to occur,  with the following dependence on metallicity: all the models studied for $Z=0.02$ experience flashes, the lowest mass proto-WD produced being $0.167\;M_{\odot}$. For $Z=0.01$ flashes occur above $0.169\;M_{\odot}$, for $Z=0.001$ the limit is $\sim\!0.22\;M_{\odot}$, while for $Z=0.0002$ the lower threshold \text{value} is $\sim\!0.26\;M_{\odot}$. When rotational mixing is included, all the threshold values are slightly higher than only diffusion is included, cf. Table~\ref{tab:flashes}. 
The upper limit for the occurrence of flashes is not as well constrained because  fewer models are calculated models in this mass range, given that the focus of this work is towards the lowest masses of 
helium WDs, which are the ELM~WDs. The obtained limits for hydrogen shell flashes agree well with those found in the literature for low metallicity \citep{serenelli01}, but at solar metallicity we obtain somewhat lower values than \cite{althaus13}. 

\begin{table}
\centering
\caption{Proto-WD mass ranges for hydrogen shell flashes. For a given model category and a given metallicity, the threshold mass for flashes also depends on the initial mass of the donor star.}
\label{tab:flashes}
\begin{tabular}{llll} 
\hline
$Z$& category& $M_{\rm{flash,min}}$ ($M_{\odot}$) & $M_{\rm{flash,max}}$ ($M_{\odot}$)\\
\hline		
0.02 &basic& 0.212  &0.305$-$0.319\\
0.02 &diffusion& <0.167  & >0.392\\
0.02 &diffusion+rotation& <0.167  & >0.321\\
\hline	    
0.01 &basic &0.222 &0.305$-$0.375 \\
0.01 &diffusion&0.167 &>0.291 \\
0.01 &diffusion+rotation& <0.181 &>0.32\\
\hline	    
0.001 &basic &0.249 & 0.349$-$0.422 \\
0.001 &diffusion & 0.223 & >0.322 \\	    
0.001 &diffusion+rotation & 0.232& >0.32 \\
\hline	    
0.0002 &basic &0.282 &0.356$-$0.441 \\
0.0002 &diffusion &0.266 &>0.311 \\
0.0002 &diffusion+rotation & 0.275 & >0.292 \\  		   
\end{tabular}
\end{table}

The change in metallicity not only affects the threshold for flashes,  but also the extent of the loops in the HR diagram. In Fig.~\ref{fig:hr_metallicity} the formation and evolution of a proto-WD with a mass of $\sim 0.28\;M_{\odot}$ is shown in the HR-diagram for all the investigated metallicities. The lower the metal content, the weaker the CNO burning, and thus the loops during the CNO flashes are markedly less extended than in  models with higher metallicity. Moreover, the number of flashes increases with decreasing metallicity: while the models for $Z=0.02$ and $Z=0.01$ 
experience just one hydrogen shell flash, the model at $Z=0.001$ goes through two flashes, and at $Z=0.0002$ the star experiences three hydrogen shell flashes. The interval of masses for which flashes occur is also affected by metallicity. For $Z=0.02$ and $Z=0.01$, a $0.28\;M_{\odot}$ helium WD is close to the upper mass limit where hydrogen shell flashes occur, while for $Z=0.001$ and $Z=0.0002$, a $0.28\;M_{\odot}$ helium WD is located close to the lower mass limit of the hydrogen shell flash interval. We stress that the number of flashes decreases with increasing  mass of the WD and varies between 0 and 7 flashes for our computed models (see  Appendix~\ref{appendix:tables}). 

\subsection{Inheritance of proto-WDs: the hydrogen envelope mass}\label{temporal_det}
\begin{figure}
\centering
\includegraphics[width=1.0\columnwidth]{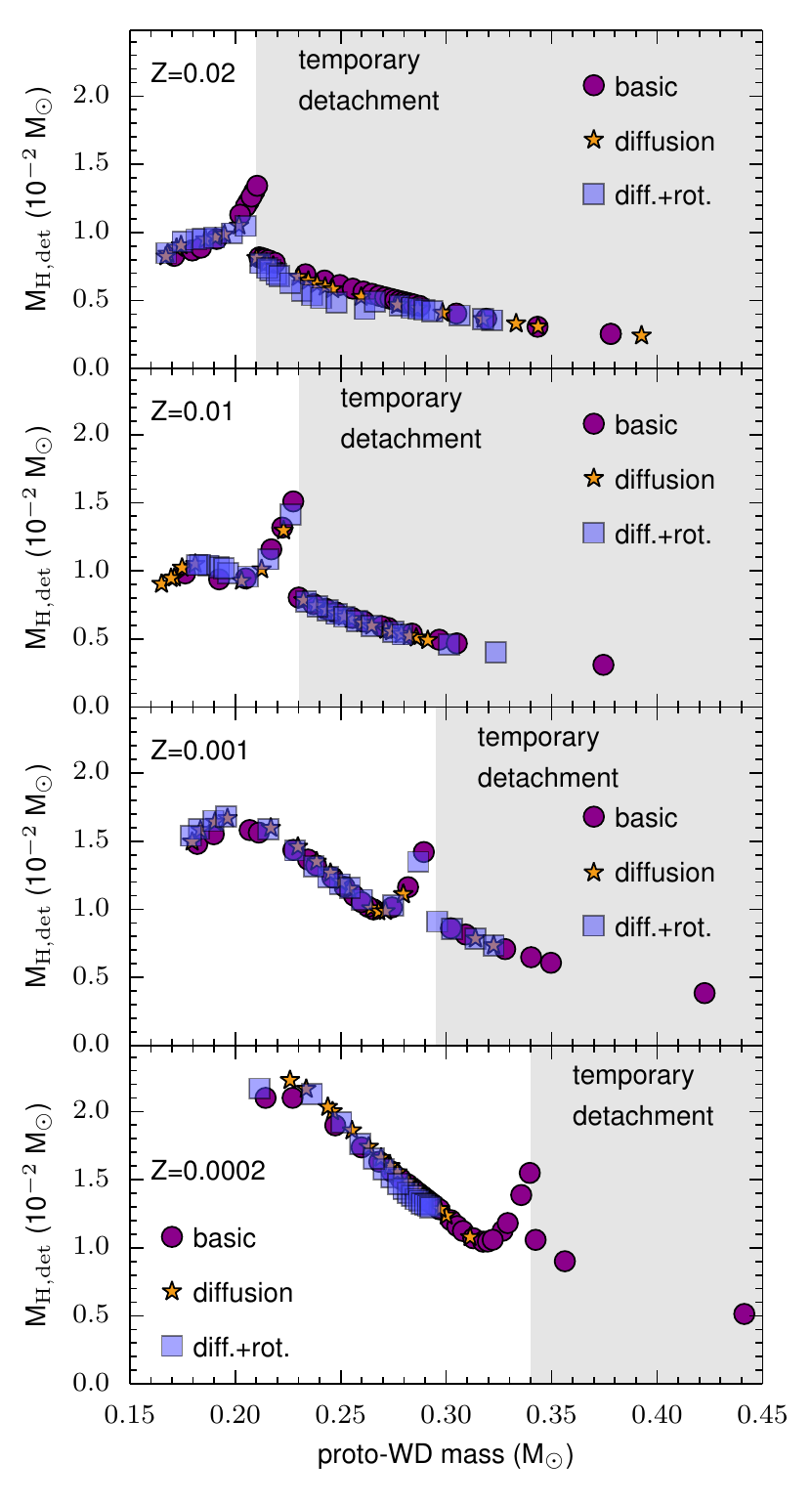}
\caption{Hydrogen envelope mass at the end of the mass-transfer phase (Roche-lobe detachment) for the basic stars (purple circles), for the stars with diffusion only (orange stars) and for the stars with diffusion+rotation (blue squares) for $Z=0.02$ (top panel), $Z=0.01$ (second panel), $Z=0.001$ (third panel) and $Z=0.0002$ (bottom panel) as a function of proto-WD mass. The grey shaded area  denotes the stars that undergo a temporary Roche-lobe detachment (see text for details).}
\label{fig:h_end_mass_transfer}
\end{figure}

Figure~\ref{fig:h_end_mass_transfer} shows the hydrogen envelope mass at the end of the mass-transfer phase (Roche-lobe detachment), $M_{\rm H, det}$, as a function of the proto-WD mass for all the computed models. For a given metallicity, the models with diffusion and with diffusion+rotation have very similar values of $M_{\rm H, det}$ as the basic models. The general trend is that the lower the mass of the proto-WD, the higher $M_{\rm H, det}$. The features in $M_{\rm H, det}$ are given by the evolutionary history of the progenitor (donor) star and depend on the point in its evolution at which mass transfer is initiated. We note a jump in the hydrogen envelope mass at $\sim\!0.21\;M_{\odot}$, $\sim\!0.23\;M_{\odot}$, $\sim\!0.29\;M_{\odot}$,  and  $\sim\!0.34\;M_{\odot}$ for $Z=0.02$, $Z=0.01$, $Z=0.001$, and $Z=0.0002$, respectively. This can be understood as discussed below. 
The shell hydrogen burning produces a convective envelope. When the convective envelope reaches its deepest extent, a hydrogen abundance gradient is produced between the region of the star mixed by the convective envelope and the layers below (which are rich in helium). When the hydrogen burning shell 
passes through this chemical discontinuity, the hydrogen burning rate drops, the radius contracts on a Kelvin-Helmholtz timescale and, as a result, the mass transfer will cease. The same  phenomenon is responsible for the occurrence of the luminosity bump in red-giant stars  \citep[e.g.][]{thomas67,christen2015}, first discussed in the context of temporary Roche-lobe detachment in LMXBs in \citet{tauris99}. 

The interruption of the mass transfer can be a temporary effect if the envelope is massive enough, such that when the burning shell has passed through the discontinuity, the star still has enough material to burn and can therefore resume its mass transfer. If its envelope has been stripped to a greater extent, then the donor star is unable to resume mass transfer and a proto-WD is formed.
This discontinuity in $M_{\rm H, det}$, observed at all the metallicities studied, distinguishes the systems that undergo this type of temporary detachment (the systems on the right-hand or upper  side of the discontinuity) from the systems in which the hydrogen shell burning passes through the 
hydrogen abundance discontinuity without being able to resume mass transfer afterwards (the systems on the left-hand or lower side of the discontinuity). This explains the increasing values of 
$M_{\rm H, det}$ with $M_{\mathrm{proto-WD}}$ just below the discontinuity. 

\subsection{$\Delta t_{\rm proto}$: the contraction timescale for proto-WDs}\label{contraction_time}

\begin{figure*}
\centering
\includegraphics[width=1.90\columnwidth]{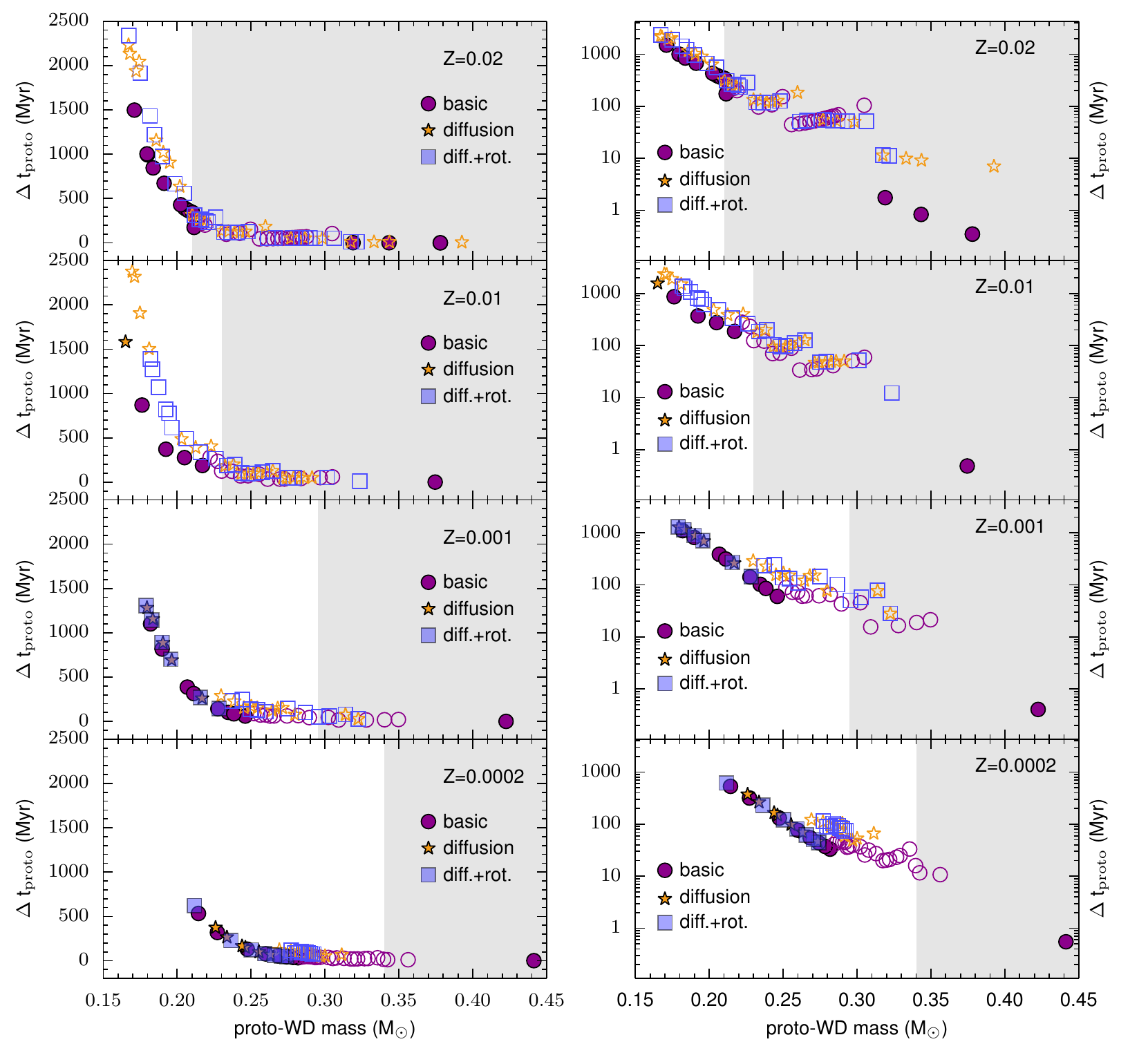}
\caption{$\Delta t_{\rm proto}$ for the basic models (purple circles), for the models with diffusion only (orange stars), and for the models with diffusion+rotation (blue squares) as a function of proto-WD mass for $Z=0.02$ (top panels), $Z=0.01$ (second panels), $Z=0.001$ (third panels) and $Z=0.0002$ (bottom panels). The left-hand panels show a linear scale in $\Delta t_{\rm proto}$ while the right-hand panels show a log~scale. Open symbols denote models that experience hydrogen shell flashes, whereas filled symbols denote models that avoid flashes. The grey shaded regions indicate temporary Roche-lobe detachment.}
\label{fig:deltat_proto}
\end{figure*}

As has been discussed earlier in this work, after the end of the LMXB mass-transfer phase, a certain amount of time, $\Delta t_{\rm proto}$, is required by the newly formed object, the proto-WD, to contract and reach its cooling track. This timescale, from Roche-lobe detachment to the beginning of the cooling track (defined as when $T_{\rm eff}$ reaches its maximum value), depends on the mass of the proto-WD and can reach up to 2~Gyr for the lowest mass proto-WDs \citep{itla14} down to $10-100\;{\rm Myr}$ for the highest mass helium WDs. More importantly, \citet{itla14} have shown that $\Delta t_{\rm proto}$ is not influenced by the occurrence of hydrogen shell flashes, and consequently, the suggested dichotomy in WD cooling times produced by hydrogen flashes was called into question.  

The determination of $\Delta t_{\rm proto}$ is important, especially  for determining the age of MSPs with helium WD companions independently of the spin-down of the MSP 
\citep[e.g.][]{kerkwijk05,antoniadis12,Bassa16}. During this phase, the proto-WD appears  to be  bloated, meaning that its radius is significantly larger than the radius of a cold WD of similar mass. As the timescale for this contraction phase ($\Delta t_{\rm proto}$) is predicted to be relatively long, a number of ELM~WDs should be observed in this bloated stage. One example suggested by \citet{itla14} is PSR~J1816+4510, based on observations by \citet{kaplan12,kaplan13}.

Figure~\ref{fig:deltat_proto} shows $\Delta t_{\rm proto}$ for all our computed models. One feature is the occurrence of clustering in the data that groups the proto-WDs that undergo the same number of flashes (see Appendix~\ref{appendix:tables}). As discussed before, the models with diffusion only and diffusion+rotation behave in a very similar way. In general, $\Delta t_{\rm proto}$ is larger  than in the basic models when diffusion is included  because of the additional flashes.  For $Z=0.02$ and $Z=0.01$ there is a smooth transition of $\Delta t_{\rm proto}$ around the limit of the occurrence of flashes (models  that experienced hydrogen shell flashes are plotted with open symbols). We recall that for these high metallicities all the models with diffusion only and diffusion+rotation (except for the model with the lowest mass at $Z=0.01$) undergo unstable burning through CNO hydrogen shell flashes. However, for $Z=0.001$ and $Z=0.0002$ we note  a slight increase in $\Delta t_{\rm proto}$ around the lowest  threshold for flashes for all models where diffusion is included. 

The maximum value of $T_{\rm eff}$ reached during the proto-WD phase, however, is very sensitive to both the time and the spatial resolution with which the stellar structure is computed. With this in mind, and taking into account that $\Delta t_{\rm proto}$ is relatively small around the lowest threshold for flashes at these low metallicities  the results shown in Fig.~\ref{fig:deltat_proto} do not present evidence for a dichotomy in $\Delta t_{\rm proto}$  that is due to hydrogen flashes. However, for the long-term evolution on the WD cooling track the situation is different, as we  discuss below.
 
\begin{figure}
\centering
\includegraphics[width=\columnwidth]{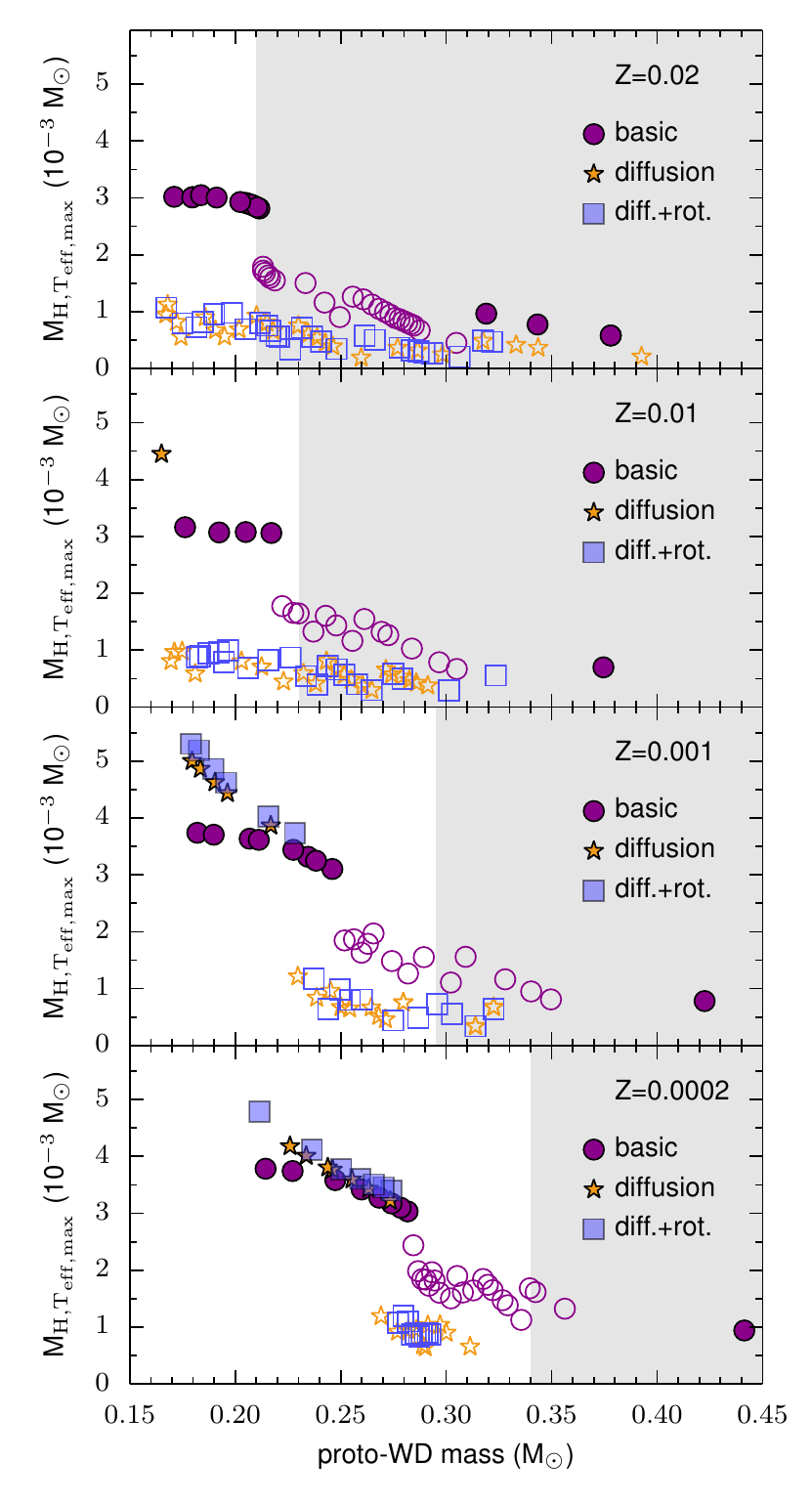}
\caption{Hydrogen envelope mass at the beginning of the cooling track (maximum $T_{\rm eff}$) for the basic models (purple circles), for the models with diffusion only (orange stars) and for the models with diffusion+rotation (blue squares)  as a function of proto-WD mass. See Fig.~\ref{fig:deltat_proto} for further explanations.}
\label{fig:h_envelope_cooling}
\end{figure}

\begin{figure}
\centering
\includegraphics[width=\columnwidth]{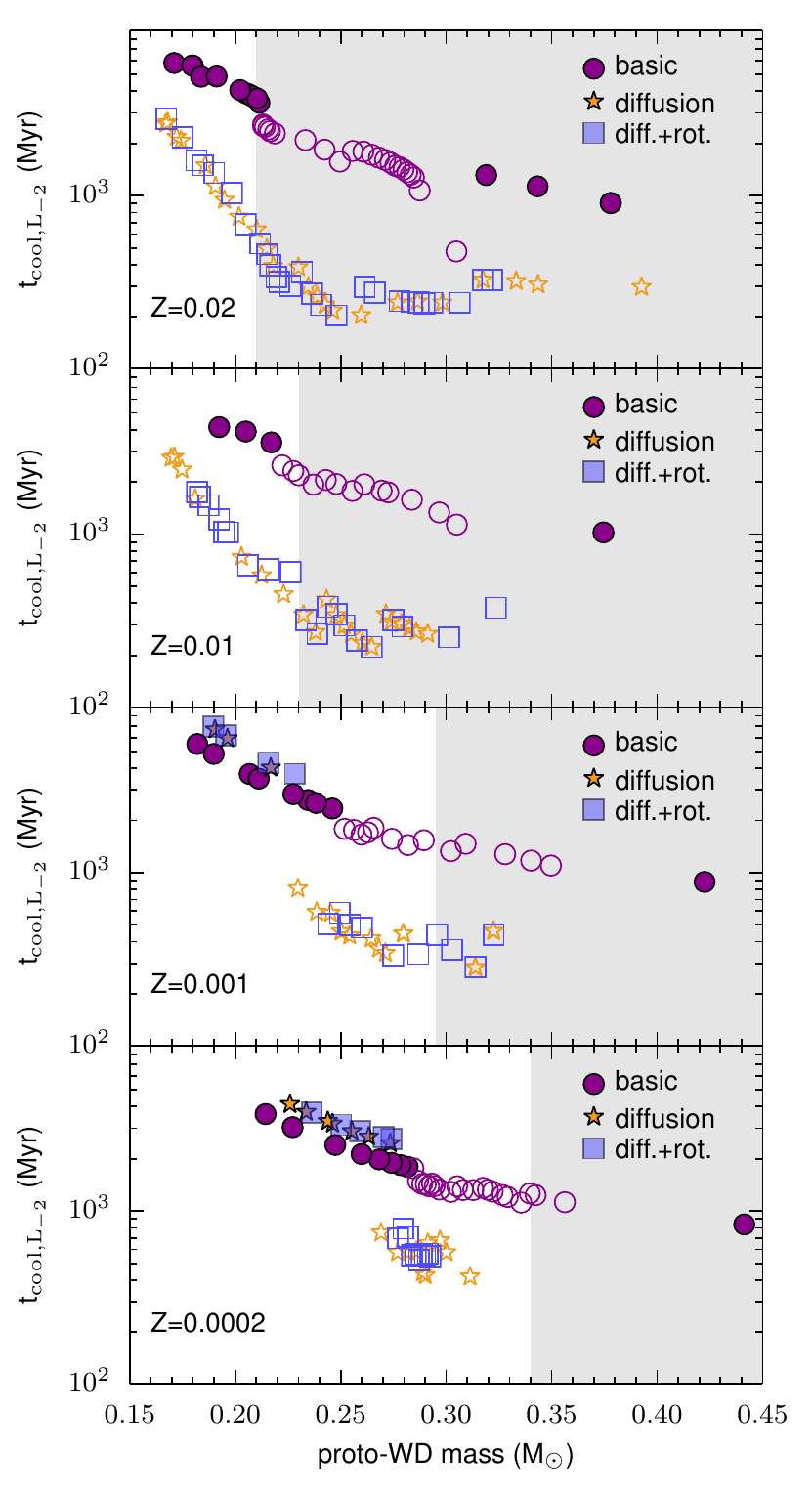}
\caption{Timescale $t_{\mathrm{cool, L_{-2}}}$ (see text) for the basic models (purple circles), for the models with diffusion only (orange stars), and for the models with diffusion+rotation (blue squares) as a function of proto-WD mass. See Fig.~\ref{fig:deltat_proto} for further explanations.}
\label{fig:cool_luminosity}
\end{figure} 

\subsection{Dichotomy  on ELM~WD cooling tracks}\label{cooling_track}
The hydrogen envelope mass is an important parameter that determines the long-term cooling timescale for WDs. Following the work of \cite{itla14}, we consider the beginning of the cooling track as the moment at which the proto-WD reaches its maximum value of $T_{\mathrm{eff}}$. Figure~\ref{fig:h_envelope_cooling} shows the remaining hydrogen envelope mass when the proto-WD reaches the maximum $T_{\rm eff}$ ($M_{\mathrm{H,T_{\mathrm{eff, max}}}}$) and finally settles on the cooling track, as  a function of the mass of the proto-WD. Again, the large scatter is related to the number of flashes (between $0-7$) that the proto-WD  experiences. At all metallicities we note a jump in $M_{\mathrm{H,T_{\mathrm{eff, max}}}}$ that occurs at the lowest threshold for flashes.
At low metallicities, the effect is more pronounced in models with diffusion and diffusion+rotation, whereas at Z=0.02  the discontinuity is only seen in the basic models (all the systems for which element diffusion is considered experience hydrogen flashes). When element diffusion is included, the hydrogen envelope mass at the beginning of the cooling track is typically twice as small as when 
element diffusion is neglected (basic models), except for models at low metallicities and masses below the flash threshold. This significantly affects the cooling times of these objects. 

For example, consider an ELM~WD with a mass of  $\sim0.20\;M_\odot$.  Figure~\ref{fig:h_envelope_cooling} shows that flashes at high metallicities ($Z=0.02$ and $Z=0.01$)
cause the amount of remaining hydrogen envelope mass at 
 $T_{\rm eff}^{\rm max}$ ($M_{\mathrm{H,T_{\mathrm{eff, max}}}}\!\sim\!1.0\!\times 10^{-3}\;M_{\odot}$) to be  up to  five times smaller than for the cases with lower metallicities ($Z=0.001$ and $Z=0.0002$) where flashes do not develop and a thick $\sim\!5.0\times 10^{-3}\;M_{\odot}$ residual hydrogen envelope remains at the onset of the cooling track. Hence, it is clear that we do see a dichotomy in the long-term cooling ages of ELM~WDs, such that those proto-WDs that  experience flashes will have thin hydrogen envelopes and therefore shorter cooling timescales, whereas proto-WDs that avoid flashes will have relatively thick hydrogen envelopes and cool on a much longer timescale. It is important to stress  that the threshold mass at which this transition occurs is dependent on metallicity.

\begin{figure}
\centering
\includegraphics[width=\columnwidth]{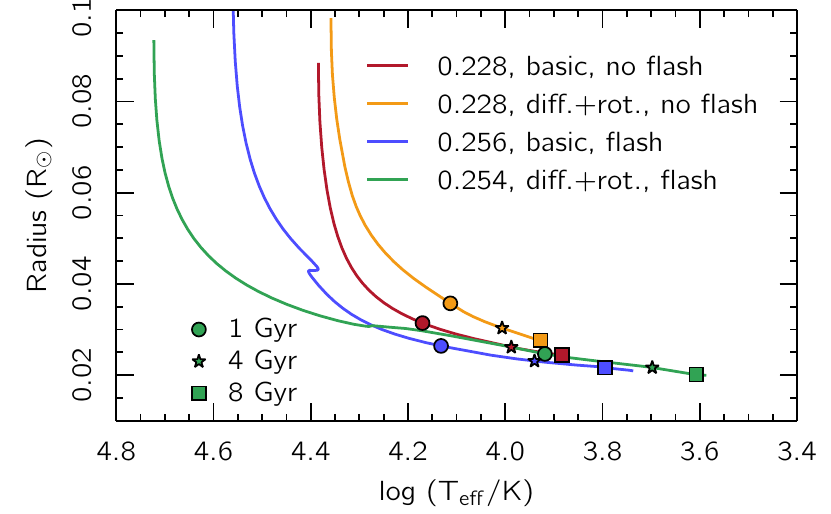}
\caption{Different cooling curves illustrating the differences between basic models (without diffusion and rotation) and models that include  element diffusion and rotation. All models have a metallicity of $Z=0.001$. The value of $M_{\rm WD}/M_{\odot}$ is shown for each track,  and the difference in masses can partly explain the different cooling rates. However, more important in this respect is the occurrence of hydrogen shell flashes (blue and green curves), which accelerates the cooling and creates a dichotomy in WD cooling ages (see text).}
\label{fig:cooling-curves}
\end{figure}

We now analyse the dichotomy in long-term cooling in more detail.
Figure~\ref{fig:cool_luminosity} shows the time from the end of the LMXB mass transfer until the WD luminosity reaches $\log (L/L_{\odot})=-2.0$, $t_{\mathrm{cool, L_{-2}}}$ (including  $\Delta t_{\rm proto}$). Some of our computed models are not plotted because  they reached an age of 14~Gyr before $\log (L/L_{\odot})=-2.0$. For the basic models, independent of metallicity, there is a small difference in cooling times between the WDs that experience flashes and those for which the hydrogen burning in the shell is stable. However, for the WDs computed with diffusion and diffusion+rotation,  the difference in cooling times between systems with and without flashes can be as large as 3~Gyr, see Fig.\ref{fig:cooling-curves}. We conclude that when element diffusion is included, the occurrence of hydrogen shell flashes does indeed produce a (metallicity-dependent) dichotomy in the cooling times of helium WDs.

Only when element diffusion is neglected in the modelling there is no or only a very small difference between the WDs that experience unstable hydrogen burning 
compared to those for which the residual hydrogen burning is stable, independent of metallicity. 
This can explain the findings of \citet{itla14}, who evolved their stellar models without element diffusion and thus questioned the dichotomy idea. 

\begin{figure}
\centering
\includegraphics[width=\columnwidth]{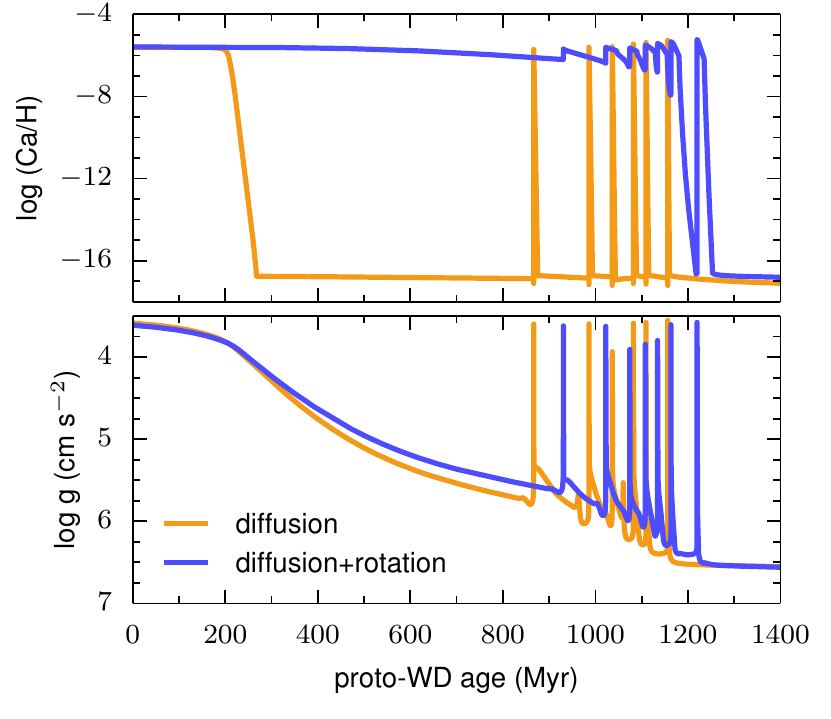}
\caption{Evolution of $\log {\rm (Ca/H)}$ at the stellar surface (top panel) and log~$g$ (bottom panel) for a $0.185\;M_{\odot}$ proto-WD computed with diffusion only (orange) and diffusion+rotation (blue) for Z=0.02. The starting point ($t=0$) is defined at the moment of Roche-lobe detachment.} 
\label{fig:metals_behaviour}
\end{figure}

%%%%%%%%%%%%%%%%%%%%%%%%%%%%%%%%%%%%%%%%%%%%%%%%%%%%%%%%%%%%%%%%%%%%%%%%%%%%%%%%%%%%%%%%%%%%%%%%%%%%%%%%%%%%
\section{Discussion}\label{discussion}
\subsection{Rotational mixing: source of surface metals?}
\begin{figure*}
\centering
\subfloat[]{\includegraphics[width=0.49\textwidth]{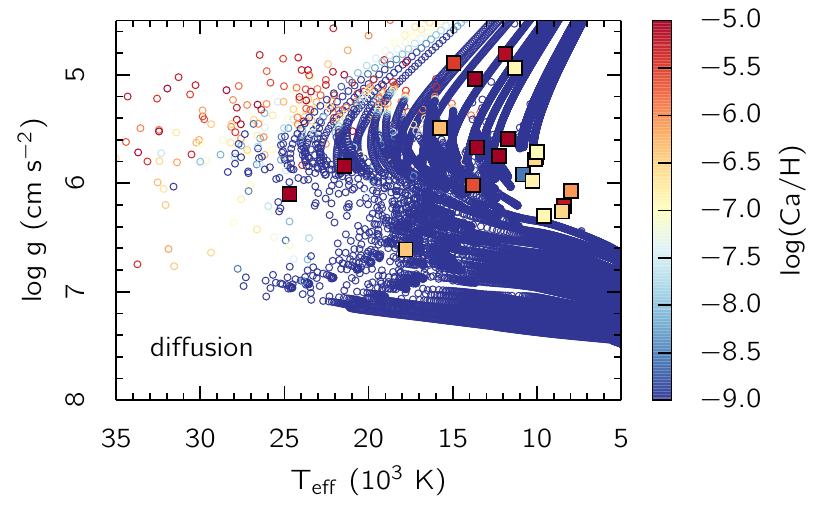}}\quad
\subfloat[]{\includegraphics[width=0.49\textwidth]{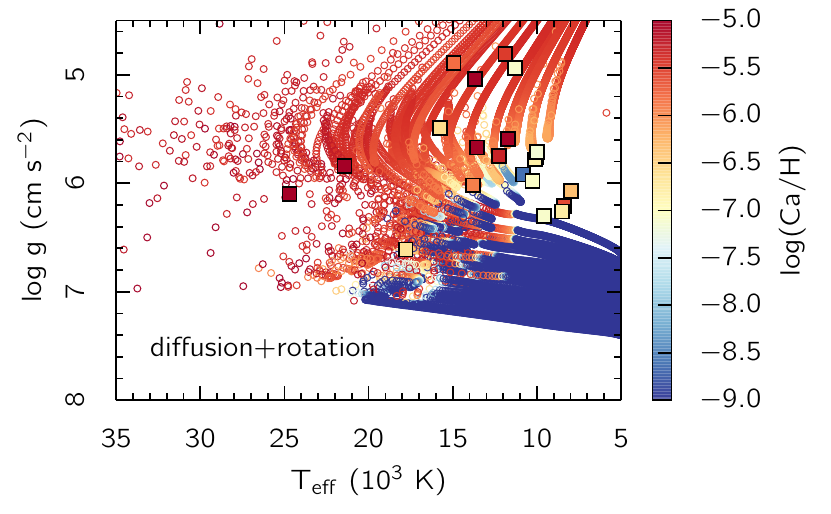}}\\
\caption{Surface gravity versus effective temperature for the models with diffusion only (a), and diffusion+rotation (b), both calculated for a metallicity of $Z=0.02$. Each evolutionary track is plotted as a dot at 0.5~Myr intervals and colour-coded according to the value of $\log~{\rm(Ca/H)}$ at the stellar surface. The over-plotted squares are the observed (proto)WDs  with measured Ca abundances taken from \citet{gianninas14}.}
\label{fig:metals}
\end{figure*}

In the past few years, metals, especially calcium, were detected in the spectra of  ELM~WDs with a surface gravity lower than $\sim$5.9. Metals sink below the atmosphere on a timescale much shorter than the evolutionary  timescale of the proto-WD, which means that another process is required to either counteract the gravitational settling or replenish the depleted metals.
There are several possible processes that can be responsible for the observed surface composition of ELM~WDs. For a detailed discussion, we refer to \cite{gianninas14} 
and \cite{hermes14b}. For higher mass (carbon-oxygen) WDs, the presence of metals in their atmosphere is explained by accretion from circumstellar debris discs formed by tidal disruption of planetary bodies \citep[e.g][]{debes2002,jura2007}, which are detectable through excess flux in the IR \citep[e.g][]{farihi09, kilic06}. This scenario seems unlikely for ELM~WDs given that their compact orbits make the existence of a debris disk dynamically difficult to explain.

\citet{kaplan13} suggested that the observed metals are brought to the surface by the pulse-driven convection developed during a hydrogen shell flash. However, shortly after the convection zone vanishes, the metals will sink below the stellar surface as a result of  gravitational settling.\\
A mechanism that can  counteract diffusion would be radiative levitation. However, as \cite{hermes14} showed, radiative levitation alone cannot explain the observed abundances, especially in the case of calcium. They suggested that in addition to radiative levitation, another support mechanism such as rotational mixing is likely required to explain the observed pattern in the metal abundances of ELM~WDs. Here, we discuss the effect of rotational mixing in determining the surface composition of ELM~WDs.

\begin{figure*}
\subfloat[]{\includegraphics[width=0.51\textwidth]{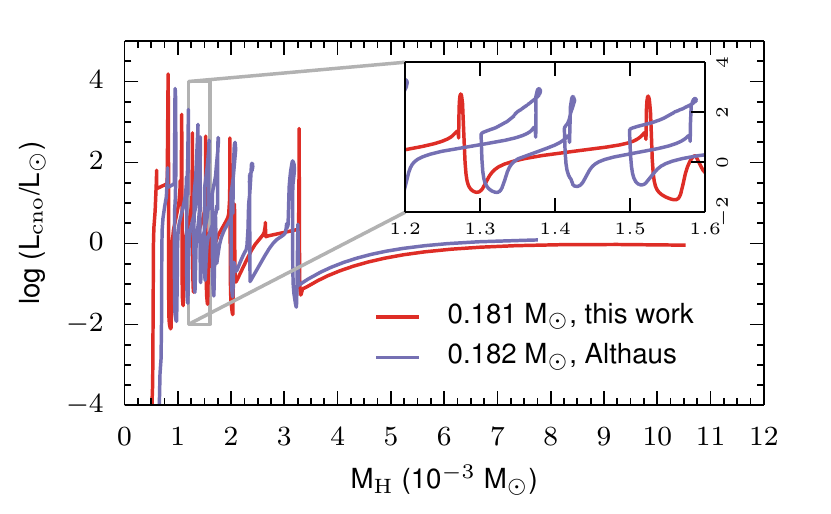}}\quad
\subfloat[]{\includegraphics[width=0.51\textwidth]{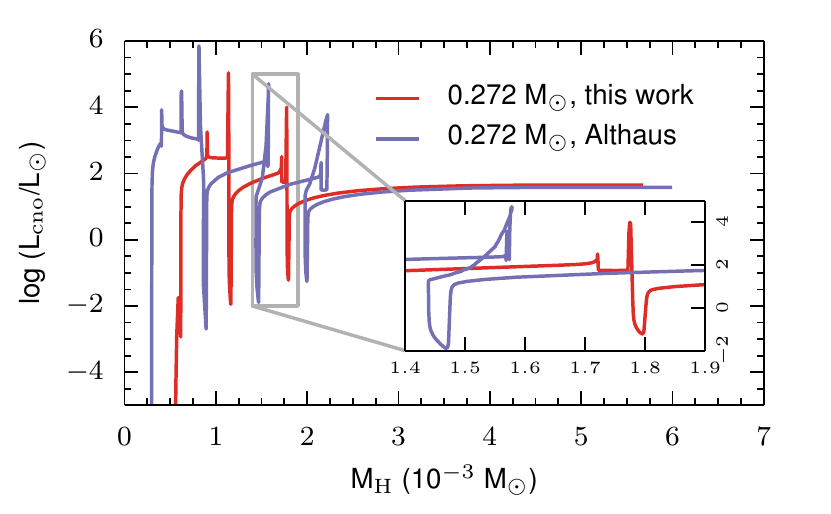}}\\
\caption{Comparison of evolutionary tracks of \citet{althaus13} and this work (including diffusion only) for a $\sim\!0.18\;M_{\odot}$~WD~(a) and  $\sim\!0.27\;M_{\odot}$~WD~(b). 
The post-RLO evolution of luminosity, given by CNO burning, is plotted as a function of the hydrogen envelope mass.  Time goes from right to left. The inset shows an artefact of hydrogen production during the shell flashes in the \citet{althaus13} models, see text.}  
\label{fig:althaus_comp_hydrogen}
\end{figure*}

\begin{figure*}
\subfloat[]{\includegraphics[width=0.5\textwidth]{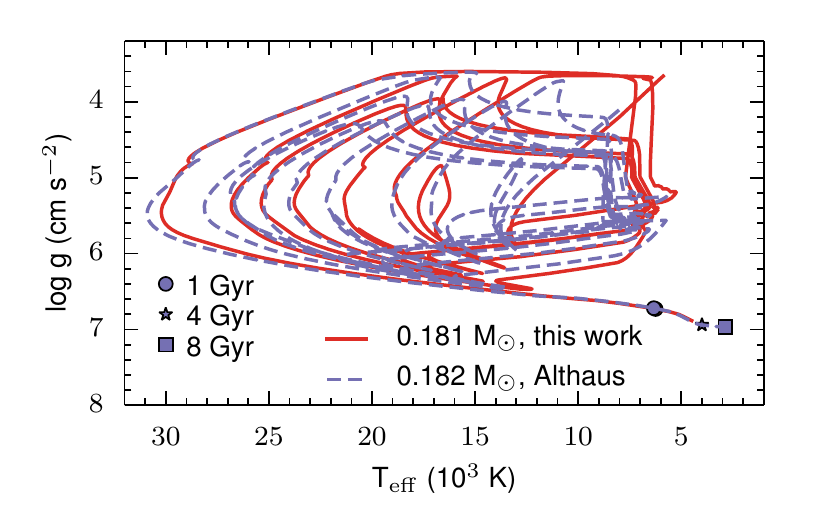}}\quad
\subfloat[]{\includegraphics[width=0.5\textwidth]{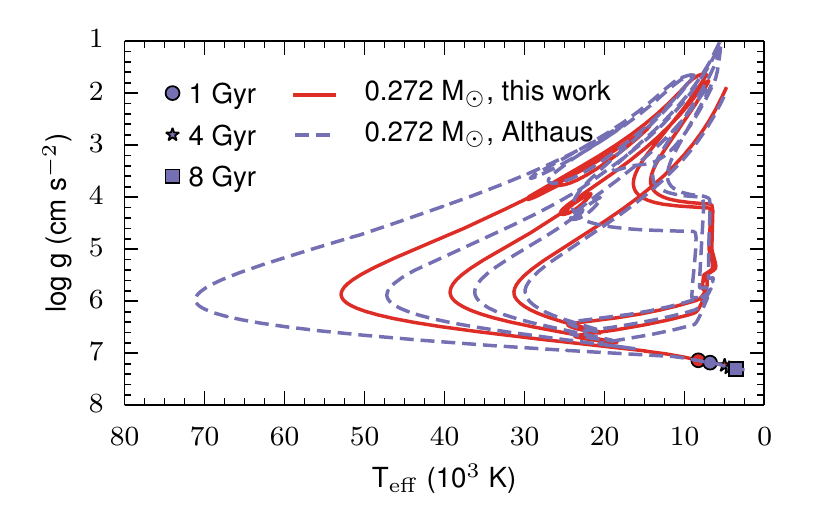}}\\
\caption{Evolutionary tracks in the ($T_{\rm eff},\log g$)--diagram for the same models as in Fig.~\ref{fig:althaus_comp_hydrogen}.  The circles, stars, and squares indicate cooling ages of 1, 4, and 8 Gyr, respectively. For the 0.18 M$_{\odot}$ WD,  the 1 Gyr symbols for the two models almost coincide.  Moreover, we note that the models from this work  are  only calculated  to an age of 14 Gyr, which prevents cooling ages exceeding 3 Gyr for the 0.18 M$_{\odot}$ model and 6 Gyr for the 0.27 M$_{\odot}$ model. }
\label{fig:althaus_comp}
\end{figure*}

Figure~\ref{fig:metals_behaviour} shows the evolution of $\log {\rm (Ca/H)}$ at the stellar surface from the beginning of the proto-WD phase (Roche-lobe detachment) to several hundred Myr onto the cooling track. In the model  that only includes diffusion calcium sinks much faster beneath the surface than the proto-WD evolutionary timescale after the mass transfer ends. It is brought back to the surface through to the pulse-driven convection zone that is developed by the occurrence of a hydrogen shell flash, only to  quickly sink again as a result of the gravitational settling. In the diffusion+rotation model, rotational mixing at the surface acts against the gravitational settling (see Sect.~\ref{rotational_mixing}). As the proto-WD advances towards higher surface gravity, rotational mixing becomes less efficient than gravitational settling. During the proto-WD phase, the star may experience several episodes of radial expansion followed by contraction  and may also develop zones of convection through the hydrogen flashes. This interplay between convection, expansion (low surface gravity), contraction (high surface gravity), and rotational mixing explains the pattern shown in Fig.~\ref{fig:metals_behaviour}. When the proto-WD enters the cooling track, the surface gravity steadily increases and gravitational settling finally  overcomes the mixing  that is due to rotation. 
As a long-term result, the metals will sink below the surface, leaving behind a pure hydrogen envelope.  

We plott in Fig.~\ref{fig:metals} all the models with $Z=0.02$ computed with diffusion only (left panel) and diffusion+rotation (right panel). The points are spaced at intervals of 0.5~Myr and 
colour-coded according to the value of $\log {\rm (Ca/H)}$. Over-plotted are the data points from \cite{gianninas14}. The left panel clearly shows that the flash scenario discussed above cannot explain the observations. On the other hand, when rotational mixing is included, we can qualitatively explain the presence of calcium in the spectra of proto-WDs as a natural result of their evolution. We recall that the observational data most likely belong to populations with different metallicities, while in Fig.~\ref{fig:metals} we only plot our models with $Z=0.02$. 
The lack of observations of proto-WDs at high T$_{\mathrm{eff}}$ arises because  the detection limit of Ca lines depends on the effective temperature (see Fig.~9 in \cite{gianninas14}).

\subsection{Comparison with previous work}\label{subsec:compare}
As discussed in \citet{gianninas15} and \citet{bours15}, the models of \cite{itla14} (from here on I14) and \cite{althaus13}(A13) show a relatively large discrepancy in their cooling ages. 
Although the initial binary parameters, and to some extent the metallicities, are different in the two sets of models, the main difference is that the models of A13 include element diffusion, which has an important role in reducing the hydrogen envelope mass through flashes (cf. Sect.~\ref{cooling_track}), and thus consequently leads to accelerated cooling and therefore younger cooling ages than in I14 models.

Here, we compare  our new models including element diffusion (but without rotational mixing to enable comparison) (I16) with the A13 models.  We also adopted the same initial binary parameters, 
namely a $1.0\;M_{\odot}$ donor star and a $1.4\;M_{\odot}$ neutron star accretor, and also applied the same metallicity of $Z=0.01$. A13 found that the lower mass limit for which hydrogen shell flashes occur is somewhere in the interval $0.176-0.182\;M_{\odot}$ (i.e. between the last model with stable shell burning and the first model that  experiences flashes), while we find a lower mass limit of $0.165-0.169\;M_{\odot}$. 

Figure~\ref{fig:althaus_comp_hydrogen} compares the I16 models with those of A13 by showing the evolution of luminosity produced by CNO burning as a function of hydrogen envelope mass for a $\sim\!0.18\;M_{\odot}$ and a $\sim\!0.27\;M_{\odot}$ (proto) helium WD. One important difference is the hydrogen envelope mass left  at Roche-lobe detachment. For the $0.18\;M_{\odot}$ WD, the model of I16 initially has a more massive hydrogen envelope, while for the $0.27\;M_{\odot}$ WD it is the opposite. We mention again that in our models diffusion acts from the ZAMS, in contrast with the A13 models, where diffusion is turned on during the proto-WD evolution. Figure~\ref{fig:althaus_comp_hydrogen} also shows that the A13 models contain a numerical artefact by which hydrogen is created during CNO burning (see inset). \\

In Fig.~\ref{fig:althaus_comp} we present a ($T_{\rm eff},\log g$)--diagram and compare the evolutionary tracks from Fig.~\ref{fig:althaus_comp_hydrogen}. The main difference are additional mass-transfer episodes in the I16 models as a result of a few vigorous flashes. This effect will in the end leave a slightly lower WD mass at the beginning of the cooling track. These differences in proto-WD evolution, combined with the artificial creation of hydrogen in the A13 models, result in slight differences on the cooling track. However, the difference between the A13 models and the I16 models are significantly smaller than when comparing the A13 models with the I14 models \citep{gianninas15,bours15}. 

\subsection{Relation of mass to orbital period in WDs}
When low-mass stars ($< 2.3\;M_{\odot}$) reach the red-giant branch, the radius of the star mainly depends on the mass of the degenerate helium core and is almost entirely independent of the mass of the envelope \citep{refsdal71,webbink83}. For the formation of binary MSPs, this relation proves to be very important because it provides a correlation between the mass of the newly formed WD and $P_{\mathrm{orb}}$ following the mass transfer episode \citep{savonije87,joss87,rappaport95,tauris99,nelson04,devito10,lin11,shao12,jia2014,itl14}.

\begin{figure}
\centering
\includegraphics[width=\columnwidth]{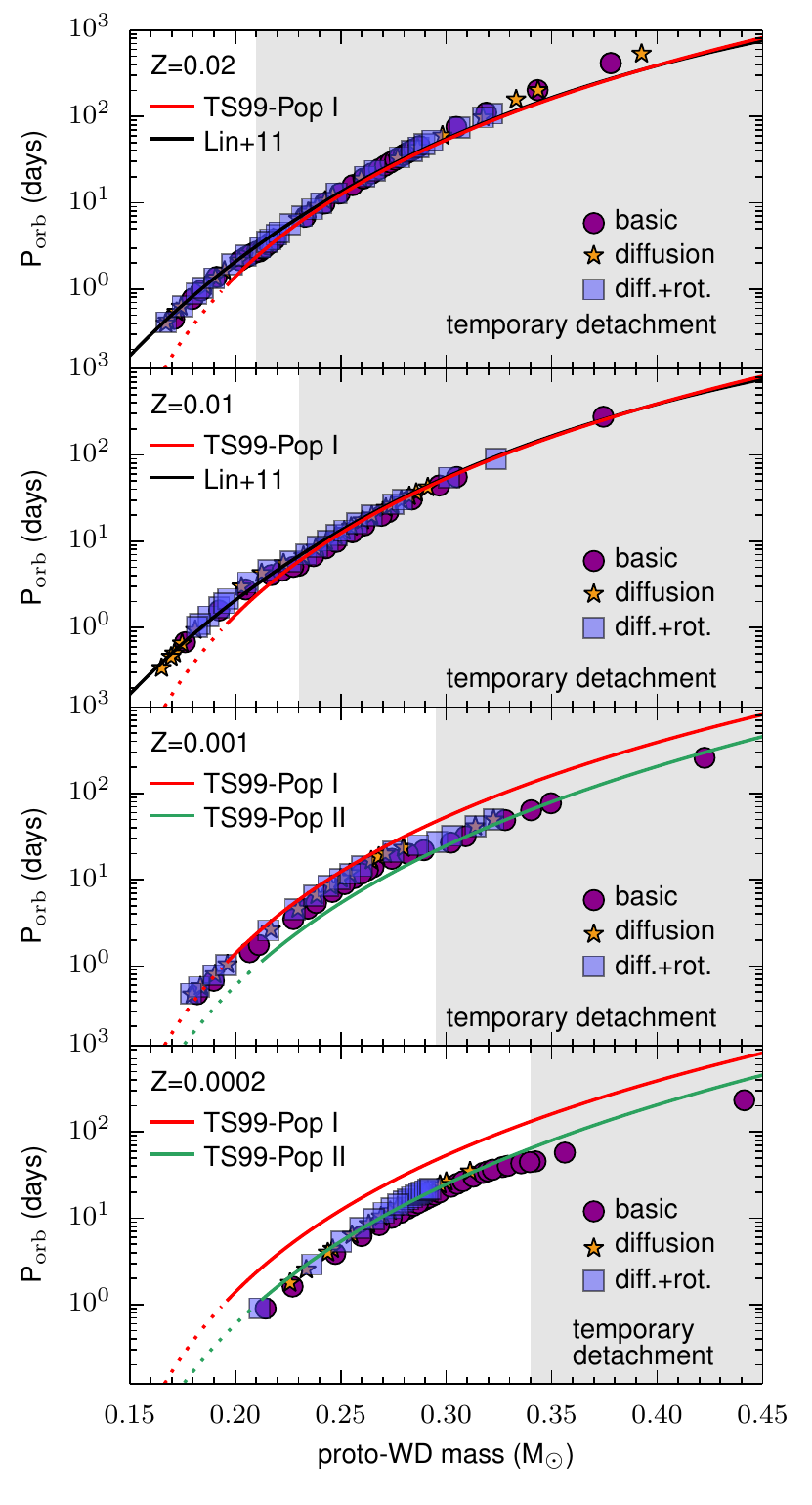}
\caption{($M_{\rm WD},P_{\rm orb}$)-relation for $Z=0.02$ (top panel), $Z=0.01$ (second panel), $Z=0.001$ (third panel), and $Z=0.0002$ (bottom panel). The grey shaded area denotes the LMXB systems that  undergo a temporary detachment (see text). The purple circles represent the basic models, the orange stars the models with diffusion only and the blue squares the models with diffusion+rotation. Over-plotted are the theoretical relations by \citet{tauris99} and \citet{lin11} for the respective metallicity.}
\label{fig:orbital_period}
\end{figure}

Figure~\ref{fig:orbital_period} shows the ($M_{\rm WD},P_{\rm orb}$)-relation for all the models computed in this work. Our results are in fine agreement with \citet{tauris99} for systems with $P_{\rm orb}>1-2\;{\rm days}$. For close-orbit systems with $P_{\rm orb}<1\;{\rm day}$
our results agree well with \citet{lin11} and \citet{itl14}. We note a slight discontinuity in the relation, which is dependent on metallicity as discussed in Sect.~\ref{temporal_det}. This weak break in the ($M_{\rm WD},P_{\rm orb}$)-relation was previously reported by other authors \citep[e.g.][]{nelson04,jia2014}. 

%%%%%%%%%%%%%%%%%%%%%%%%%%%%%%%%%%%%%%%%%%%%%%%%%%%%%%%%%%%%%%%%%%%%%%%%%%%%%%%%%%%%%%%%%%%%%%%%%%%%%%%%%%%%
\section{Conclusions}\label{conclusions}
We computed a grid of models for ELM~WDs with different metallicities. For each metallicity, we computed three types of models with different physics included: 
i) basic models (with no element diffusion nor rotation), ii) diffusion (including element diffusion), and iii) diffusion+rotation. For the first time, we took into account the combined effects 
of rotational mixing and element diffusion in the evolution of WD progenitors and during the proto-WD phase and WD cooling. The main results obtained are summarized as follows:
\begin{enumerate}[(i)]
\item We confirm that element diffusion plays a significant role in the evolution of proto-WDs that experience hydrogen shell flashes. We also confirm that the unstable burning is triggered by the diffusive hydrogen tail reaching the hot deep layers inside the star \citep{althaus2001}. The formation of the hydrogen tail is a cyclic process and depends on the available hydrogen in the envelope. Consequently, the number of flashes experienced by a proto-WD of a given mass is increased, leading to reduced hydrogen envelope mass and subsequent accelerated cooling, compared with the models without element diffusion.

\item Rotational mixing counteracts the effect of gravitational settling in the surface layers of young bloated ELM~proto-WDs, but its efficiency is reduced towards the end of the proto-WD phase,  
when the star contracts and its surface gravity increases. As a consequence, our new evolutionary models including rotational mixing  predict that ELM proto-WDs have mixed H/He envelopes during a significant part of their evolution before settling on the cooling track, in accordance with recent observational evidence from pulsations in ELM proto-WDs \citep{gianninas2016}. Except for this, the general properties, such as the number of flashes, are not strongly influenced  by the presence of rotational mixing.  

\item Although the hydrogen envelope left after detachment from the LMXB phase is a small fraction of the total  WD mass, it has a very high angular momentum content compared to the core. Because the proto-WD contracts while this hydrogen is burnt and angular momentum mixes inwards following each flash, the resulting WD on the cooling track is spun up significantly  to a  rotation period well below the orbital period.

\item The hydrogen envelope mass in newborn proto-WDs is influenced by the evolutionary stage of the donor star at the moment when LMXB mass transfer is initiated. In particular, we found that LMXB donor stars that experience a temporary contraction will produce proto-WDs with a significantly reduced hydrogen envelope mass at the moment of the final Roche-lobe detachment. In general, the shorter the orbital period at the onset of the LMXB~phase, the lower the mass of the proto-WD and the higher the final envelope mass. The hydrogen envelope mass is also metallicity dependent, such that (for the same proto-WD mass) the lower the metallicity, the higher the envelope mass. 
\item In general, our resulting mass range for the occurrence of flashes is similar to those found in the literature. For $Z=0.02$, all our models with diffusion experience flashes,  and for $Z=0.01$ we obtain a lower limit of $\sim\!0.16\;M_{\odot}$, compared to $\sim\!0.18\;M_{\odot}$ found by \cite{althaus13}.

\item We identified two timescales relevant for  understanding the evolution of (proto) WDs. The evolutionary timescale of the  contraction of proto-WDs, $\Delta t_{\rm proto}$, can reach up to 2.5~Gyr for the lowest proto-WD masses, while for the higher WD masses it is $<100\;{\rm Myr}$. When element diffusion is included, this timescale is slightly increased by more numerous flashes than for  the case when element diffusion is neglected. As concluded in \cite{itla14}, we did not find a dichotomy 
in the $\Delta t_{\rm proto}$ distribution with respect to the occurrence of flashes, but rather a smooth transition. However,  we note a small increase in $\Delta t_{\rm proto}$ close to the flash limit for $Z=0.001$ and $Z=0.002$.\\ 
The cooling timescale $t_{\rm cool,L_{-2}}$ describes the evolution of the WD  from LMXB detachment until the the WD has evolved well down the cooling track and reaches log ($L/L_{\odot})=-2$. This timescale is mostly determined by the hydrogen envelope mass  that remains after the proto-WD phase. In the basic models (without element diffusion or rotation), independent of metallicity, there is a very small difference between the models that experience flashes and those that do not. However, we confirm, as first stated by \cite{althaus2001}, that the situation is different when element diffusion is included, which leads to a dichotomy in the cooling timescale of ELM WDs. 

\item We investigated whether the observed metal features in ELM proto-WDs might be linked to the internal evolution of these objects. In particular, we analysed the evolution of the surface abundance of calcium and concluded that rotational mixing is a key component for producing the observed pattern in the ($\log g,T_{\rm eff}$)--diagram. 
\end{enumerate}
The systematic investigation presented here for the effects of thermal and chemical diffusion, 
gravitational settling, and rotational mixing on a wide range of LMXBs with different initial masses, orbital periods, and metallicity,  leads us to conclude that theoretical models must include these aspects when  they are compared to observational data of ELM~WDs. Hence, with these improved models at hand, the implications are better constraints on the true ages and masses of these WDs and therefore also on the ages and masses of their companion stars, such as millisecond radio pulsars.  Moreover, the grid of models we presented might be further used for  astroseismology calculations to determine the pulsational behaviour of ELM proto-WDs and ELM WDs.  
\section*{Acknowledgements}
We thank the anonymous referee for helpful advice. A.G.I. thanks Alejandra Romero and Elvijs Matrozis for very useful discussions and BWin Technologies Ltd. for the computing cores on which the simulations in this work were performed.  

\bibliographystyle{aa}
\bibliography{paper_IV_references} 

\begin{thebibliography}{110}
\expandafter\ifx\csname natexlab\endcsname\relax\def\natexlab#1{#1}\fi

\bibitem[{{Althaus} \& {Benvenuto}(2000)}]{althaus2000}
{Althaus}, L.~G. \& {Benvenuto}, O.~G. 2000, \mnras, 317, 952

\bibitem[{{Althaus} {et~al.}(2013){Althaus}, {Miller Bertolami}, \&
  {C{\'o}rsico}}]{althaus13}
{Althaus}, L.~G., {Miller Bertolami}, M.~M., \& {C{\'o}rsico}, A.~H. 2013,
  \aap, 557, A19

\bibitem[{{Althaus} {et~al.}(2009){Althaus}, {Panei}, {Romero}, {Rohrmann},
  {C{\'o}rsico}, {Garc{\'{\i}}a-Berro}, \& {Miller Bertolami}}]{althaus2009}
{Althaus}, L.~G., {Panei}, J.~A., {Romero}, A.~D., {et~al.} 2009, \aap, 502,
  207

\bibitem[{{Althaus} {et~al.}(2001{\natexlab{a}}){Althaus}, {Serenelli}, \&
  {Benvenuto}}]{althaus2001}
{Althaus}, L.~G., {Serenelli}, A.~M., \& {Benvenuto}, O.~G. 2001{\natexlab{a}},
  \mnras, 323, 471

\bibitem[{{Althaus} {et~al.}(2001{\natexlab{b}}){Althaus}, {Serenelli}, \&
  {Benvenuto}}]{althaus2001b}
{Althaus}, L.~G., {Serenelli}, A.~M., \& {Benvenuto}, O.~G. 2001{\natexlab{b}},
  \apj, 554, 1110

\bibitem[{{Althaus} {et~al.}(2001{\natexlab{c}}){Althaus}, {Serenelli}, \&
  {Benvenuto}}]{althaus2001c}
{Althaus}, L.~G., {Serenelli}, A.~M., \& {Benvenuto}, O.~G. 2001{\natexlab{c}},
  \mnras, 324, 617

\bibitem[{{Amaro-Seoane} {et~al.}(2012){Amaro-Seoane}, {Aoudia}, {Babak},
  {Bin{\'e}truy}, {Berti}, {Boh{\'e}}, {Caprini}, {Colpi}, {Cornish},
  {Danzmann}, {Dufaux}, {Gair}, {Jennrich}, {Jetzer}, {Klein}, {Lang}, {Lobo},
  {Littenberg}, {McWilliams}, {Nelemans}, {Petiteau}, {Porter}, {Schutz},
  {Sesana}, {Stebbins}, {Sumner}, {Vallisneri}, {Vitale}, {Volonteri}, \&
  {Ward}}]{amaro2012}
{Amaro-Seoane}, P., {Aoudia}, S., {Babak}, S., {et~al.} 2012, Classical and
  Quantum Gravity, 29, 124016

\bibitem[{{Andrews} {et~al.}(2014){Andrews}, {Price-Whelan}, \&
  {Ag{\"u}eros}}]{andrews14}
{Andrews}, J.~J., {Price-Whelan}, A.~M., \& {Ag{\"u}eros}, M.~A. 2014, \apjl,
  797, L32

\bibitem[{{Antoniadis} {et~al.}(2012){Antoniadis}, {van Kerkwijk}, {Koester},
  {Freire}, {Wex}, {Tauris}, {Kramer}, \& {Bassa}}]{antoniadis12}
{Antoniadis}, J., {van Kerkwijk}, M.~H., {Koester}, D., {et~al.} 2012, \mnras,
  423, 3316

\bibitem[{{Bassa} {et~al.}(2016){Bassa}, {Antoniadis}, {Camilo}, {Cognard},
  {Koester}, {Kramer}, {Ransom}, \& {Stappers}}]{Bassa16}
{Bassa}, C.~G., {Antoniadis}, J., {Camilo}, F., {et~al.} 2016, \mnras, 455,
  3806

\bibitem[{{Bildsten} {et~al.}(2007){Bildsten}, {Shen}, {Weinberg}, \&
  {Nelemans}}]{bilsten07}
{Bildsten}, L., {Shen}, K.~J., {Weinberg}, N.~N., \& {Nelemans}, G. 2007,
  \apjl, 662, L95

\bibitem[{{Bours} {et~al.}(2015){Bours}, {Marsh}, {G{\"a}nsicke}, {Tauris},
  {Istrate}, {Badenes}, {Dhillon}, {Gal-Yam}, {Hermes}, {Kengkriangkrai},
  {Kilic}, {Koester}, {Mullally}, {Prasert}, {Steeghs}, {Thompson}, \&
  {Thorstensen}}]{bours15}
{Bours}, M.~C.~P., {Marsh}, T.~R., {G{\"a}nsicke}, B.~T., {et~al.} 2015,
  \mnras, 450, 3966

\bibitem[{{Brown} {et~al.}(2016){Brown}, {Gianninas}, {Kilic}, {Kenyon}, \&
  {Allende Prieto}}]{warren2016}
{Brown}, W.~R., {Gianninas}, A., {Kilic}, M., {Kenyon}, S.~J., \& {Allende
  Prieto}, C. 2016, \apj, 818, 155

\bibitem[{{Brown} {et~al.}(2013){Brown}, {Kilic}, {Allende Prieto},
  {Gianninas}, \& {Kenyon}}]{elm5}
{Brown}, W.~R., {Kilic}, M., {Allende Prieto}, C., {Gianninas}, A., \&
  {Kenyon}, S.~J. 2013, \apj, 769, 66

\bibitem[{{Brown} {et~al.}(2010){Brown}, {Kilic}, {Allende Prieto}, \&
  {Kenyon}}]{elm1}
{Brown}, W.~R., {Kilic}, M., {Allende Prieto}, C., \& {Kenyon}, S.~J. 2010,
  \apj, 723, 1072

\bibitem[{{Brown} {et~al.}(2012){Brown}, {Kilic}, {Allende Prieto}, \&
  {Kenyon}}]{elm3}
{Brown}, W.~R., {Kilic}, M., {Allende Prieto}, C., \& {Kenyon}, S.~J. 2012,
  \apj, 744, 142

\bibitem[{{Cadelano} {et~al.}(2015){Cadelano}, {Pallanca}, {Ferraro},
  {Salaris}, {Dalessandro}, {Lanzoni}, \& {Freire}}]{cadelano15}
{Cadelano}, M., {Pallanca}, C., {Ferraro}, F.~R., {et~al.} 2015, \apj, 812, 63

\bibitem[{{Camilo} {et~al.}(1994){Camilo}, {Thorsett}, \&
  {Kulkarni}}]{camilo94}
{Camilo}, F., {Thorsett}, S.~E., \& {Kulkarni}, S.~R. 1994, \apjl, 421, L15

\bibitem[{{Cassisi} {et~al.}(2007){Cassisi}, {Potekhin}, {Pietrinferni},
  {Catelan}, \& {Salaris}}]{cassisi07}
{Cassisi}, S., {Potekhin}, A.~Y., {Pietrinferni}, A., {Catelan}, M., \&
  {Salaris}, M. 2007, \apj, 661, 1094

\bibitem[{{Christensen-Dalsgaard}(2015)}]{christen2015}
{Christensen-Dalsgaard}, J. 2015, \mnras, 453, 666

\bibitem[{{Clayton}(2013)}]{clayton2013}
{Clayton}, G.~C. 2013, in Astronomical Society of the Pacific Conference
  Series, Vol. 469, 18th European White Dwarf Workshop., 133

\bibitem[{{C{\'o}rsico} \& {Althaus}(2014{\natexlab{a}})}]{corsico14}
{C{\'o}rsico}, A.~H. \& {Althaus}, L.~G. 2014{\natexlab{a}}, \aap, 569, A106

\bibitem[{{C{\'o}rsico} \& {Althaus}(2014{\natexlab{b}})}]{corsico14b}
{C{\'o}rsico}, A.~H. \& {Althaus}, L.~G. 2014{\natexlab{b}}, \apjl, 793, L17

\bibitem[{{C{\'o}rsico} {et~al.}(2016){C{\'o}rsico}, {Althaus}, {Serenelli},
  {Kepler}, {Jeffery}, \& {Corti}}]{corsico16}
{C{\'o}rsico}, A.~H., {Althaus}, L.~G., {Serenelli}, A.~M., {et~al.} 2016,
  \aap, 588, A74

\bibitem[{{Corti} {et~al.}(2016){Corti}, {Kanaan}, {C{\'o}rsico}, {Kepler},
  {Althaus}, {Koester}, \& {S{\'a}nchez Arias}}]{corti2016}
{Corti}, M.~A., {Kanaan}, A., {C{\'o}rsico}, A.~H., {et~al.} 2016, \aap, 587,
  L5

\bibitem[{{De Vito} \& {Benvenuto}(2010)}]{devito10}
{De Vito}, M.~A. \& {Benvenuto}, O.~G. 2010, \mnras, 401, 2552

\bibitem[{{Debes} \& {Sigurdsson}(2002)}]{debes2002}
{Debes}, J.~H. \& {Sigurdsson}, S. 2002, \apj, 572, 556

\bibitem[{{Detmers} {et~al.}(2008){Detmers}, {Langer}, {Podsiadlowski}, \&
  {Izzard}}]{detmes2008}
{Detmers}, R.~G., {Langer}, N., {Podsiadlowski}, P., \& {Izzard}, R.~G. 2008,
  \aap, 484, 831

\bibitem[{{Driebe} {et~al.}(1998){Driebe}, {Schoenberner}, {Bloecker}, \&
  {Herwig}}]{dribe98}
{Driebe}, T., {Schoenberner}, D., {Bloecker}, T., \& {Herwig}, F. 1998, \aap,
  339, 123

\bibitem[{{Farihi} {et~al.}(2009){Farihi}, {Jura}, \& {Zuckerman}}]{farihi09}
{Farihi}, J., {Jura}, M., \& {Zuckerman}, B. 2009, \apj, 694, 805

\bibitem[{{Ferguson} {et~al.}(2005){Ferguson}, {Alexander}, {Allard}, {Barman},
  {Bodnarik}, {Hauschildt}, {Heffner-Wong}, \& {Tamanai}}]{ferguson05}
{Ferguson}, J.~W., {Alexander}, D.~R., {Allard}, F., {et~al.} 2005, \apj, 623,
  585

\bibitem[{{Foley}(2015)}]{foley2015}
{Foley}, R.~J. 2015, \mnras, 452, 2463

\bibitem[{{Fontaine} \& {Michaud}(1979)}]{fontaine79}
{Fontaine}, G. \& {Michaud}, G. 1979, \apj, 231, 826

\bibitem[{{Gautschy}(2013)}]{gautschy13}
{Gautschy}, A. 2013, ArXiv e-prints

\bibitem[{{Gianninas} {et~al.}(2016){Gianninas}, {Curd}, {Fontaine}, {Brown},
  \& {Kilic}}]{gianninas2016}
{Gianninas}, A., {Curd}, B., {Fontaine}, G., {Brown}, W.~R., \& {Kilic}, M.
  2016, ArXiv e-prints

\bibitem[{{Gianninas} {et~al.}(2014{\natexlab{a}}){Gianninas}, {Dufour},
  {Kilic}, {Brown}, {Bergeron}, \& {Hermes}}]{gianninas14}
{Gianninas}, A., {Dufour}, P., {Kilic}, M., {et~al.} 2014{\natexlab{a}}, \apj,
  794, 35

\bibitem[{{Gianninas} {et~al.}(2014{\natexlab{b}}){Gianninas}, {Hermes},
  {Brown}, {Dufour}, {Barber}, {Kilic}, {Kenyon}, \& {Harrold}}]{gianninas14b}
{Gianninas}, A., {Hermes}, J.~J., {Brown}, W.~R., {et~al.} 2014{\natexlab{b}},
  \apj, 781, 104

\bibitem[{{Gianninas} {et~al.}(2015){Gianninas}, {Kilic}, {Brown}, {Canton}, \&
  {Kenyon}}]{gianninas15}
{Gianninas}, A., {Kilic}, M., {Brown}, W.~R., {Canton}, P., \& {Kenyon}, S.~J.
  2015, \apj, 812, 167

\bibitem[{{Grevesse} \& {Sauval}(1998)}]{grevesse98}
{Grevesse}, N. \& {Sauval}, A.~J. 1998, \ssr, 85, 161

\bibitem[{{Heber}(2016)}]{heber2016}
{Heber}, U. 2016, ArXiv e-prints

\bibitem[{{Heger} {et~al.}(2000){Heger}, {Langer}, \& {Woosley}}]{heger00}
{Heger}, A., {Langer}, N., \& {Woosley}, S.~E. 2000, \apj, 528, 368

\bibitem[{{Heger} {et~al.}(2005){Heger}, {Woosley}, \& {Spruit}}]{heger05}
{Heger}, A., {Woosley}, S.~E., \& {Spruit}, H.~C. 2005, \apj, 626, 350

\bibitem[{{Henyey} {et~al.}(1965){Henyey}, {Vardya}, \&
  {Bodenheimer}}]{henyey65}
{Henyey}, L., {Vardya}, M.~S., \& {Bodenheimer}, P. 1965, \apj, 142, 841

\bibitem[{{Hermes} {et~al.}(2014{\natexlab{a}}){Hermes}, {Brown}, {Kilic},
  {Gianninas}, {Chote}, {Sullivan}, {Winget}, {Bell}, {Falcon}, {Winget},
  {Mason}, {Harrold}, \& {Montgomery}}]{hermes14}
{Hermes}, J.~J., {Brown}, W.~R., {Kilic}, M., {et~al.} 2014{\natexlab{a}},
  \apj, 792, 39

\bibitem[{{Hermes} {et~al.}(2014{\natexlab{b}}){Hermes}, {G{\"a}nsicke},
  {Koester}, {Bours}, {Townsley}, {Farihi}, {Marsh}, {Littlefair}, {Dhillon},
  {Gianninas}, {Breedt}, \& {Raddi}}]{hermes14b}
{Hermes}, J.~J., {G{\"a}nsicke}, B.~T., {Koester}, D., {et~al.}
  2014{\natexlab{b}}, \mnras, 444, 1674

\bibitem[{{Hermes} {et~al.}(2012{\natexlab{a}}){Hermes}, {Kilic}, {Brown},
  {Winget}, {Allende Prieto}, {Gianninas}, {Mukadam}, {Cabrera-Lavers}, \&
  {Kenyon}}]{hermes2012b}
{Hermes}, J.~J., {Kilic}, M., {Brown}, W.~R., {et~al.} 2012{\natexlab{a}},
  \apjl, 757, L21

\bibitem[{{Hermes} {et~al.}(2013{\natexlab{a}}){Hermes}, {Montgomery},
  {Gianninas}, {Winget}, {Brown}, {Harrold}, {Bell}, {Kenyon}, {Kilic}, \&
  {Castanheira}}]{hermes13b}
{Hermes}, J.~J., {Montgomery}, M.~H., {Gianninas}, A., {et~al.}
  2013{\natexlab{a}}, \mnras, 436, 3573

\bibitem[{{Hermes} {et~al.}(2013{\natexlab{b}}){Hermes}, {Montgomery},
  {Winget}, {Brown}, {Gianninas}, {Kilic}, {Kenyon}, {Bell}, \&
  {Harrold}}]{hermes13}
{Hermes}, J.~J., {Montgomery}, M.~H., {Winget}, D.~E., {et~al.}
  2013{\natexlab{b}}, \apj, 765, 102

\bibitem[{{Hermes} {et~al.}(2012{\natexlab{b}}){Hermes}, {Montgomery},
  {Winget}, {Brown}, {Kilic}, \& {Kenyon}}]{hermes12}
{Hermes}, J.~J., {Montgomery}, M.~H., {Winget}, D.~E., {et~al.}
  2012{\natexlab{b}}, \apjl, 750, L28

\bibitem[{{Hu} {et~al.}(2011){Hu}, {Tout}, {Glebbeek}, \& {Dupret}}]{hu11}
{Hu}, H., {Tout}, C.~A., {Glebbeek}, E., \& {Dupret}, M.-A. 2011, \mnras, 418,
  195

\bibitem[{{Hurley} {et~al.}(2002){Hurley}, {Tout}, \& {Pols}}]{hurley2002}
{Hurley}, J.~R., {Tout}, C.~A., \& {Pols}, O.~R. 2002, \mnras, 329, 897

\bibitem[{{Iben} \& {MacDonald}(1985)}]{iben85}
{Iben}, Jr., I. \& {MacDonald}, J. 1985, \apj, 296, 540

\bibitem[{{Iben} \& {Tutukov}(1984)}]{iben84}
{Iben}, Jr., I. \& {Tutukov}, A.~V. 1984, \apjs, 54, 335

\bibitem[{{Iglesias} \& {Rogers}(1993)}]{iglesias93}
{Iglesias}, C.~A. \& {Rogers}, F.~J. 1993, \apj, 412, 752

\bibitem[{{Iglesias} \& {Rogers}(1996)}]{iglesias96}
{Iglesias}, C.~A. \& {Rogers}, F.~J. 1996, \apj, 464, 943

\bibitem[{{Istrate} {et~al.}(2014{\natexlab{a}}){Istrate}, {Tauris}, \&
  {Langer}}]{itl14}
{Istrate}, A.~G., {Tauris}, T.~M., \& {Langer}, N. 2014{\natexlab{a}}, \aap,
  571, A45

\bibitem[{{Istrate} {et~al.}(2014{\natexlab{b}}){Istrate}, {Tauris}, {Langer},
  \& {Antoniadis}}]{itla14}
{Istrate}, A.~G., {Tauris}, T.~M., {Langer}, N., \& {Antoniadis}, J.
  2014{\natexlab{b}}, \aap, 571, L3

\bibitem[{{Itoh} {et~al.}(1987){Itoh}, {Kohyama}, \& {Takeuchi}}]{itoh87}
{Itoh}, N., {Kohyama}, Y., \& {Takeuchi}, H. 1987, \apj, 317, 733

\bibitem[{{Jeffery} \& {Saio}(2013)}]{jeffsaio2013}
{Jeffery}, C.~S. \& {Saio}, H. 2013, \mnras, 435, 885

\bibitem[{{Jia} \& {Li}(2014)}]{jia2014}
{Jia}, K. \& {Li}, X.-D. 2014, \apj, 791, 127

\bibitem[{{Joss} {et~al.}(1987){Joss}, {Rappaport}, \& {Lewis}}]{joss87}
{Joss}, P.~C., {Rappaport}, S., \& {Lewis}, W. 1987, \apj, 319, 180

\bibitem[{{Jura} {et~al.}(2007){Jura}, {Farihi}, \& {Zuckerman}}]{jura2007}
{Jura}, M., {Farihi}, J., \& {Zuckerman}, B. 2007, \apj, 663, 1285

\bibitem[{{Kaplan} {et~al.}(2013){Kaplan}, {Bhalerao}, {van Kerkwijk},
  {Koester}, {Kulkarni}, \& {Stovall}}]{kaplan13}
{Kaplan}, D.~L., {Bhalerao}, V.~B., {van Kerkwijk}, M.~H., {et~al.} 2013, \apj,
  765, 158

\bibitem[{{Kaplan} {et~al.}(2012){Kaplan}, {Stovall}, {Ransom}, {Roberts},
  {Kotulla}, {Archibald}, {Biwer}, {Boyles}, {Dartez}, {Day}, {Ford}, {Garcia},
  {Hessels}, {Jenet}, {Karako}, {Kaspi}, {Kondratiev}, {Lorimer}, {Lynch},
  {McLaughlin}, {Rohr}, {Siemens}, {Stairs}, \& {van Leeuwen}}]{kaplan12}
{Kaplan}, D.~L., {Stovall}, K., {Ransom}, S.~M., {et~al.} 2012, \apj, 753, 174

\bibitem[{{Kepler} {et~al.}(2016){Kepler}, {Pelisoli}, {Koester}, {Ourique},
  {Romero}, {Reindl}, {Kleinman}, {Eisenstein}, {Valois}, \&
  {Amaral}}]{kepler15}
{Kepler}, S.~O., {Pelisoli}, I., {Koester}, D., {et~al.} 2016, \mnras, 455,
  3413

\bibitem[{{Kilic} {et~al.}(2011){Kilic}, {Brown}, {Allende Prieto},
  {Ag{\"u}eros}, {Heinke}, \& {Kenyon}}]{elm2}
{Kilic}, M., {Brown}, W.~R., {Allende Prieto}, C., {et~al.} 2011, \apj, 727, 3

\bibitem[{{Kilic} {et~al.}(2012){Kilic}, {Brown}, {Allende Prieto}, {Kenyon},
  {Heinke}, {Ag{\"u}eros}, \& {Kleinman}}]{elm4}
{Kilic}, M., {Brown}, W.~R., {Allende Prieto}, C., {et~al.} 2012, \apj, 751,
  141

\bibitem[{{Kilic} {et~al.}(2013){Kilic}, {Brown}, \& {Hermes}}]{kilic2013a}
{Kilic}, M., {Brown}, W.~R., \& {Hermes}, J.~J. 2013, in Astronomical Society
  of the Pacific Conference Series, Vol. 467, 9th LISA Symposium, ed.
  G.~{Auger}, P.~{Bin{\'e}truy}, \& E.~{Plagnol}, 47

\bibitem[{{Kilic} {et~al.}(2015){Kilic}, {Hermes}, {Gianninas}, \&
  {Brown}}]{kilic15}
{Kilic}, M., {Hermes}, J.~J., {Gianninas}, A., \& {Brown}, W.~R. 2015, \mnras,
  446, L26

\bibitem[{{Kilic} {et~al.}(2014){Kilic}, {Hermes}, {Gianninas}, {Brown},
  {Heinke}, {Ag{\"u}eros}, {Chote}, {Sullivan}, {Bell}, \&
  {Harrold}}]{kilic14am}
{Kilic}, M., {Hermes}, J.~J., {Gianninas}, A., {et~al.} 2014, \mnras, 438, L26

\bibitem[{{Kilic} {et~al.}(2007){Kilic}, {Stanek}, \& {Pinsonneault}}]{kilic07}
{Kilic}, M., {Stanek}, K.~Z., \& {Pinsonneault}, M.~H. 2007, \apj, 671, 761

\bibitem[{{Kilic} {et~al.}(2006){Kilic}, {von Hippel}, {Leggett}, \&
  {Winget}}]{kilic06}
{Kilic}, M., {von Hippel}, T., {Leggett}, S.~K., \& {Winget}, D.~E. 2006, \apj,
  646, 474

\bibitem[{{Koester} {et~al.}(2009){Koester}, {Voss}, {Napiwotzki},
  {Christlieb}, {Homeier}, {Lisker}, {Reimers}, \& {Heber}}]{koester09}
{Koester}, D., {Voss}, B., {Napiwotzki}, R., {et~al.} 2009, \aap, 505, 441

\bibitem[{{Latour} {et~al.}(2016){Latour}, {Heber}, {Irrgang}, {Schaffenroth},
  {Geier}, {Hillebrandt}, {R{\"o}pke}, {Taubenberger}, {Kromer}, \&
  {Fink}}]{latour15}
{Latour}, M., {Heber}, U., {Irrgang}, A., {et~al.} 2016, \aap, 585, A115

\bibitem[{{Lin} {et~al.}(2011){Lin}, {Rappaport}, {Podsiadlowski}, {Nelson},
  {Paxton}, \& {Todorov}}]{lin11}
{Lin}, J., {Rappaport}, S., {Podsiadlowski}, P., {et~al.} 2011, \apj, 732, 70

\bibitem[{{Lorimer} {et~al.}(1995){Lorimer}, {Lyne}, {Festin}, \&
  {Nicastro}}]{lorimer95}
{Lorimer}, D.~R., {Lyne}, A.~G., {Festin}, L., \& {Nicastro}, L. 1995, \nat,
  376, 393

\bibitem[{{Marsh} {et~al.}(1995){Marsh}, {Dhillon}, \& {Duck}}]{marsh95}
{Marsh}, T.~R., {Dhillon}, V.~S., \& {Duck}, S.~R. 1995, \mnras, 275, 828

\bibitem[{{Maxted} {et~al.}(2011){Maxted}, {Anderson}, {Burleigh}, {Collier
  Cameron}, {Heber}, {G{\"a}nsicke}, {Geier}, {Kupfer}, {Marsh}, {Nelemans},
  {O'Toole}, {{\O}stensen}, {Smalley}, \& {West}}]{maxted11}
{Maxted}, P.~F.~L., {Anderson}, D.~R., {Burleigh}, M.~R., {et~al.} 2011,
  \mnras, 418, 1156

\bibitem[{{Maxted} {et~al.}(2014{\natexlab{a}}){Maxted}, {Bloemen}, {Heber},
  {Geier}, {Wheatley}, {Marsh}, {Breedt}, {Sebastian}, {Faillace}, {Owen},
  {Pulley}, {Smith}, {Kolb}, {Haswell}, {Southworth}, {Anderson}, {Smalley},
  {Collier Cameron}, {Hebb}, {Simpson}, {West}, {Bochinski}, {Busuttil}, \&
  {Hadigal}}]{maxted14}
{Maxted}, P.~F.~L., {Bloemen}, S., {Heber}, U., {et~al.} 2014{\natexlab{a}},
  \mnras, 437, 1681

\bibitem[{{Maxted} {et~al.}(2014{\natexlab{b}}){Maxted}, {Serenelli}, {Marsh},
  {Catal{\'a}n}, {Mahtani}, \& {Dhillon}}]{maxted2014}
{Maxted}, P.~F.~L., {Serenelli}, A.~M., {Marsh}, T.~R., {et~al.}
  2014{\natexlab{b}}, \mnras, 444, 208

\bibitem[{{Maxted} {et~al.}(2013){Maxted}, {Serenelli}, {Miglio}, {Marsh},
  {Heber}, {Dhillon}, {Littlefair}, {Copperwheat}, {Smalley}, {Breedt}, \&
  {Schaffenroth}}]{maxted2013}
{Maxted}, P.~F.~L., {Serenelli}, A.~M., {Miglio}, A., {et~al.} 2013, \nat, 498,
  463

\bibitem[{{Nandez} {et~al.}(2015){Nandez}, {Ivanova}, \& {Lombardi}}]{nandez15}
{Nandez}, J.~L.~A., {Ivanova}, N., \& {Lombardi}, J.~C. 2015, \mnras, 450, L39

\bibitem[{{Nelemans} \& {Tauris}(1998)}]{nt98}
{Nelemans}, G. \& {Tauris}, T.~M. 1998, \aap, 335, L85

\bibitem[{{Nelson} {et~al.}(2004){Nelson}, {Dubeau}, \&
  {MacCannell}}]{nelson04}
{Nelson}, L.~A., {Dubeau}, E., \& {MacCannell}, K.~A. 2004, \apj, 616, 1124

\bibitem[{{Panei} {et~al.}(2007){Panei}, {Althaus}, {Chen}, \& {Han}}]{panei07}
{Panei}, J.~A., {Althaus}, L.~G., {Chen}, X., \& {Han}, Z. 2007, \mnras, 382,
  779

\bibitem[{{Paquette} {et~al.}(1986){Paquette}, {Pelletier}, {Fontaine}, \&
  {Michaud}}]{paquette1986}
{Paquette}, C., {Pelletier}, C., {Fontaine}, G., \& {Michaud}, G. 1986, \apjs,
  61, 197

\bibitem[{{Paxton} {et~al.}(2011){Paxton}, {Bildsten}, {Dotter}, {Herwig},
  {Lesaffre}, \& {Timmes}}]{mesa10}
{Paxton}, B., {Bildsten}, L., {Dotter}, A., {et~al.} 2011, \apjs, 192, 3

\bibitem[{{Paxton} {et~al.}(2013){Paxton}, {Cantiello}, {Arras}, {Bildsten},
  {Brown}, {Dotter}, {Mankovich}, {Montgomery}, {Stello}, {Timmes}, \&
  {Townsend}}]{mesa13}
{Paxton}, B., {Cantiello}, M., {Arras}, P., {et~al.} 2013, \apjs, 208, 4

\bibitem[{{Paxton} {et~al.}(2015){Paxton}, {Marchant}, {Schwab}, {Bauer},
  {Bildsten}, {Cantiello}, {Dessart}, {Farmer}, {Hu}, {Langer}, {Townsend},
  {Townsley}, \& {Timmes}}]{mesa3}
{Paxton}, B., {Marchant}, P., {Schwab}, J., {et~al.} 2015, \apjs, 220, 15

\bibitem[{{Perets} {et~al.}(2010){Perets}, {Gal-Yam}, {Mazzali}, {Arnett},
  {Kagan}, {Filippenko}, {Li}, {Arcavi}, {Cenko}, {Fox}, {Leonard}, {Moon},
  {Sand}, {Soderberg}, {Anderson}, {James}, {Foley}, {Ganeshalingam}, {Ofek},
  {Bildsten}, {Nelemans}, {Shen}, {Weinberg}, {Metzger}, {Piro}, {Quataert},
  {Kiewe}, \& {Poznanski}}]{perets2010}
{Perets}, H.~B., {Gal-Yam}, A., {Mazzali}, P.~A., {et~al.} 2010, \nat, 465, 322

\bibitem[{{Rappaport} {et~al.}(1995){Rappaport}, {Podsiadlowski}, {Joss}, {Di
  Stefano}, \& {Han}}]{rappaport95}
{Rappaport}, S., {Podsiadlowski}, P., {Joss}, P.~C., {Di Stefano}, R., \&
  {Han}, Z. 1995, \mnras, 273, 731

\bibitem[{{Refsdal} \& {Weigert}(1971)}]{refsdal71}
{Refsdal}, S. \& {Weigert}, A. 1971, \aap, 13, 367

\bibitem[{{Rivera-Sandoval} {et~al.}(2015){Rivera-Sandoval}, {van den Berg},
  {Heinke}, {Cohn}, {Lugger}, {Freire}, {Anderson}, {Serenelli}, {Althaus},
  {Cool}, {Grindlay}, {Edmonds}, {Wijnands}, \& {Ivanova}}]{rivera15}
{Rivera-Sandoval}, L.~E., {van den Berg}, M., {Heinke}, C.~O., {et~al.} 2015,
  \mnras, 453, 2707

\bibitem[{{Rohrmann} {et~al.}(2012){Rohrmann}, {Althaus},
  {Garc{\'{\i}}a-Berro}, {C{\'o}rsico}, \& {Miller Bertolami}}]{rohrmann12}
{Rohrmann}, R.~D., {Althaus}, L.~G., {Garc{\'{\i}}a-Berro}, E., {C{\'o}rsico},
  A.~H., \& {Miller Bertolami}, M.~M. 2012, \aap, 546, A119

\bibitem[{{Sarna} {et~al.}(2000){Sarna}, {Ergma}, \& {Ger{\v s}kevit{\v
  s}-Antipova}}]{sarna2000}
{Sarna}, M.~J., {Ergma}, E., \& {Ger{\v s}kevit{\v s}-Antipova}, J. 2000,
  \mnras, 316, 84

\bibitem[{{Savonije}(1987)}]{savonije87}
{Savonije}, G.~J. 1987, \nat, 325, 416

\bibitem[{{Serenelli} {et~al.}(2001){Serenelli}, {Althaus}, {Rohrmann}, \&
  {Benvenuto}}]{serenelli01}
{Serenelli}, A.~M., {Althaus}, L.~G., {Rohrmann}, R.~D., \& {Benvenuto}, O.~G.
  2001, \mnras, 325, 607

\bibitem[{{Serenelli} {et~al.}(2002){Serenelli}, {Althaus}, {Rohrmann}, \&
  {Benvenuto}}]{sarb2002}
{Serenelli}, A.~M., {Althaus}, L.~G., {Rohrmann}, R.~D., \& {Benvenuto}, O.~G.
  2002, \mnras, 337, 1091

\bibitem[{{Shao} \& {Li}(2012)}]{shao12}
{Shao}, Y. \& {Li}, X.-D. 2012, \apj, 756, 85

\bibitem[{{Solheim}(2010)}]{solheim2010}
{Solheim}, J.-E. 2010, \pasp, 122, 1133

\bibitem[{{Spruit}(2002)}]{spruit2002}
{Spruit}, H.~C. 2002, \aap, 381, 923

\bibitem[{{Tauris}(2012)}]{tauris12}
{Tauris}, T.~M. 2012, Science, 335, 561

\bibitem[{{Tauris} {et~al.}(2012){Tauris}, {Langer}, \& {Kramer}}]{tlk12}
{Tauris}, T.~M., {Langer}, N., \& {Kramer}, M. 2012, \mnras, 425, 1601

\bibitem[{{Tauris} \& {Savonije}(1999)}]{tauris99}
{Tauris}, T.~M. \& {Savonije}, G.~J. 1999, \aap, 350, 928

\bibitem[{{Thomas}(1967)}]{thomas67}
{Thomas}, H.-C. 1967, \zap, 67, 420

\bibitem[{{Thoul} {et~al.}(1994){Thoul}, {Bahcall}, \& {Loeb}}]{thoul94}
{Thoul}, A.~A., {Bahcall}, J.~N., \& {Loeb}, A. 1994, \apj, 421, 828

\bibitem[{{Van Grootel} {et~al.}(2013){Van Grootel}, {Fontaine}, {Brassard}, \&
  {Dupret}}]{grootel2013}
{Van Grootel}, V., {Fontaine}, G., {Brassard}, P., \& {Dupret}, M.-A. 2013,
  \apj, 762, 57

\bibitem[{{van Kerkwijk} {et~al.}(2005){van Kerkwijk}, {Bassa}, {Jacoby}, \&
  {Jonker}}]{kerkwijk05}
{van Kerkwijk}, M.~H., {Bassa}, C.~G., {Jacoby}, B.~A., \& {Jonker}, P.~G.
  2005, in Astronomical Society of the Pacific Conference Series, Vol. 328,
  Binary Radio Pulsars, ed. F.~A. {Rasio} \& I.~H. {Stairs}, 357

\bibitem[{{Vauclair} {et~al.}(1979){Vauclair}, {Vauclair}, \&
  {Greenstein}}]{vauclair79}
{Vauclair}, G., {Vauclair}, S., \& {Greenstein}, J.~L. 1979, \aap, 80, 79

\bibitem[{{Webbink} {et~al.}(1983){Webbink}, {Rappaport}, \&
  {Savonije}}]{webbink83}
{Webbink}, R.~F., {Rappaport}, S., \& {Savonije}, G.~J. 1983, \apj, 270, 678

\end{thebibliography}

\appendix
\section{Properties of the computed models}\label{appendix:tables}
 \begin{table*}
 \center
 \caption{Properties of selected models. The quantities given are (Cols. 1-7):
                    the (proto) WD mass, the number of hydrogen shell flashes, the orbital period and the hydrogen envelope mass at Roche-lobe detachment, 
                    the hydrogen envelope mass at maximum $T_{\rm eff}$, $\Delta t_{\rm proto}$, and the time interval (cooling timescale) from Roche-lobe detachment 
                    until the WD reaches a luminosity of $\log (L/L_{\odot})=-2$.}. 
 \begin{tabular}{lllllll}
 \hline  & & & & & & \\ 
 \multicolumn{7}{c}{$Z=0.02$, basic models (no element diffusion nor rotation)}\\ 
  & & & & & & \\ 
 \hline
 Mass ($M_{\odot}$)& $\#$ flashes& $P_{\rm orb,det}$ (d) & $M_{\rm H,det}$ ($10^{-2}\;M_{\odot}$) & $M_{\rm H,T_{\rm eff,max}}$ ($10^{-3}\;M_{\odot}$) & $\Delta t_{\rm proto}$ (Myr) & $t_{\rm cool,L_{-2}}$ (Myr) \\
 \hline
 \hline  
0.171 &    0 &      0.451 &      0.827 &      3.02 &       1500 &      5820 \\
0.180 &    0 &      0.767 &      0.867 &      3.01 &       1000 &      5630 \\
0.180 &    0 &      0.779 &      0.869 &      3.01 &       993 &      5620 \\
0.184 &    0 &      0.961 &      0.887 &      3.05 &       846 &      4850 \\
0.191 &    0 &      1.38 &      0.952 &      3.01 &       672 &      4860 \\
0.202 &    0 &      2.14 &      1.13 &      2.93 &       430 &      4060 \\
0.205 &    0 &      2.34 &      1.19 &      2.91 &       392 &      3900 \\
0.206 &    0 &      2.43 &      1.23 &      2.90 &       377 &      3850 \\
0.208 &    0 &      2.53 &      1.26 &      2.88 &       364 &      3760 \\
0.209 &    0 &      2.63 &      1.30 &      2.86 &       352 &      3690 \\
0.210 &    0 &      2.73 &      1.34 &      2.83 &       342 &      3630 \\
0.211 &    0 &      2.71 &      0.817 &      2.81 &       174 &      3430 \\
0.213 &    3 &      2.94 &      0.808 &      1.72 &       274 &      2530 \\
0.213 &    3 &      2.94 &      0.808 &      1.79 &       268 &      2570 \\
0.214 &    3 &      3.05 &      0.804 &      1.68 &       261 &      2490 \\
0.216 &    3 &      3.28 &      0.794 &      1.64 &       237 &      2410 \\
0.216 &    3 &      3.40 &      0.790 &      1.59 &       221 &      2350 \\
0.219 &    3 &      3.76 &      0.777 &      1.55 &       200 &      2280 \\
0.233 &    2 &      6.89 &      0.692 &      1.50 &       97.1 &      2090 \\
0.242 &    2 &      9.81 &      0.646 &      1.16 &       107 &      1840 \\
0.250 &    2 &      12.8 &      0.614 &      0.905 &       153 &      1570 \\
0.256 &    1 &      16.0 &      0.587 &      1.26 &       44.6 &      1810 \\
0.261 &    1 &      19.0 &      0.566 &      1.22 &       46.1 &      1800 \\
0.265 &    1 &      21.9 &      0.549 &      1.13 &       48.0 &      1720 \\
0.268 &    1 &      24.5 &      0.534 &      1.05 &       50.1 &      1660 \\
0.271 &    1 &      26.9 &      0.523 &      0.996 &       52.2 &      1610 \\
0.274 &    1 &      29.3 &      0.513 &      0.945 &       54.3 &      1550 \\
0.276 &    1 &      31.5 &      0.504 &      0.889 &       56.3 &      1490 \\
0.278 &    1 &      33.7 &      0.496 &      0.868 &       58.2 &      1460 \\
0.280 &    1 &      35.8 &      0.489 &      0.835 &       60.1 &      1410 \\
0.282 &    1 &      37.8 &      0.482 &      0.805 &       61.9 &      1360 \\
0.283 &    1 &      39.8 &      0.476 &      0.779 &       63.7 &      1320 \\
0.285 &    1 &      41.7 &      0.470 &      0.755 &       65.4 &      1270 \\
0.287 &    1 &      45.5 &      0.460 &      0.670 &       68.9 &      1070 \\
0.305 &    1 &      75.6 &      0.404 &      0.451 &       105 &      477 \\
0.319 &    0 &      111 &      0.365 &      0.965 &       1.76 &      1310 \\
0.343 &    0 &      202 &      0.307 &      0.775 &       0.838 &      1130 \\
0.378 &    0 &      414 &      0.255 &      0.581 &       0.351 &      907 \\
 \noalign{\smallskip}
 \noalign{\smallskip}
 \end{tabular} 
 \begin{flushleft}
 \end{flushleft}
 \vspace{0.7cm}
 \label{table:table1} 
 \end{table*}
\clearpage

\begin{table*}
 \center
 \caption{Properties of selected models. See Table \ref{table:table1} for a description of the parameters. }
 \begin{tabular}{lllllll}
 \hline  & & & & & & \\ 
 \multicolumn{7}{c}{$Z=0.02$, diffusion}\\ 
 & & & & & & \\
 \hline
 Mass ($M_{\odot}$)& $\#$ flashes& $P_{\rm orb,det}$ (d) & $M_{\rm H,det}$ ($10^{-2}\;M_{\odot}$) & $M_{\rm H,T_{\rm eff,max}}$ ($10^{-3}\;M_{\odot}$) & $\Delta t_{\rm proto}$ (Myr) & $t_{\rm cool,L_{-2}}$ (Myr) \\
 \hline
 \hline  
0.167 &    24 &      0.401 &      0.823 &      0.951 &       2230 &      2610 \\
0.168 &    26 &      0.431 &      0.843 &      1.12 &       2130 &      2650 \\
0.172 &    12 &      0.541 &      0.885 &      0.818 &       1940 &      2190 \\
0.174 &    10 &      0.623 &      0.904 &      0.557 &       2050 &      2080 \\
0.186 &    6 &      1.08 &      0.948 &      0.899 &       1160 &      1500 \\
0.191 &    6 &      1.35 &      0.963 &      0.674 &       1020 &      1130 \\
0.195 &    6 &      1.63 &      0.984 &      0.563 &       907 &      945 \\
0.202 &    5 &      2.14 &      1.05 &      0.689 &       636 &      755 \\
0.210 &    4 &      2.72 &      0.811 &      0.930 &       316 &      638 \\
0.215 &    4 &      3.40 &      0.775 &      0.806 &       264 &      495 \\
0.218 &    5 &      3.89 &      0.753 &      0.662 &       256 &      392 \\
0.230 &    3 &      6.45 &      0.672 &      0.753 &       134 &      387 \\
0.235 &    3 &      7.79 &      0.644 &      0.606 &       129 &      299 \\
0.239 &    4 &      9.16 &      0.620 &      0.540 &       121 &      263 \\
0.243 &    3 &      10.6 &      0.600 &      0.448 &       123 &      234 \\
0.246 &    4 &      12.0 &      0.582 &      0.390 &       128 &      218 \\
0.260 &    3 &      19.0 &      0.526 &      0.191 &       184 &      205 \\
0.277 &    3 &      32.6 &      0.467 &      0.365 &       54.9 &      242 \\
0.286 &    3 &      43.2 &      0.441 &      0.316 &       52.1 &      246 \\
0.298 &    2 &      60.0 &      0.409 &      0.246 &       50.7 &      242 \\
0.317 &    1 &      97.5 &      0.363 &      0.487 &       11.5 &      328 \\
0.333 &    2 &      157 &      0.331 &      0.416 &       9.96 &      322 \\
0.344 &    2 &      201 &      0.307 &      0.366 &       9.17 &      308 \\
0.393 &    2 &      532 &      0.242 &      0.211 &       7.03 &      298 \\
 \noalign{\smallskip}
 \noalign{\smallskip}
 \end{tabular} 
 \begin{flushleft}
 \end{flushleft}
 \vspace{0.7cm}
 \end{table*}

\begin{table*}
 \center
 \caption{Properties of selected models. See Table \ref{table:table1} for a description of the parameters.} 
 \begin{tabular}{lllllll}
 \hline  & & & & & & \\ 
 \multicolumn{7}{c}{$Z=0.02$, diffusion+rotation}\\ 
 & & & & & & \\ 
 \hline
 Mass ($M_{\odot}$)& $\#$ flashes& $P_{\rm orb,det}$ (d) & $M_{\rm H,det}$ ($10^{-2}\;M_{\odot}$) & $M_{\rm H,T_{\rm eff,max}}$ ($10^{-3}\;M_{\odot}$) & $\Delta t_{\rm proto}$ (Myr) & $t_{\rm cool,L_{-2}}$ (Myr) \\
 \hline
 \hline  
0.167 &    23 &      0.412 &      0.848 &      1.07 &       2340 &      2770 \\
0.175 &    10 &      0.628 &      0.929 &      0.794 &       1910 &      2170 \\
0.182 &    8 &      0.886 &      0.948 &      0.725 &       1430 &      1600 \\
0.185 &    7 &      1.03 &      0.954 &      0.821 &       1220 &      1500 \\
0.190 &    6 &      1.33 &      0.963 &      0.956 &       976 &      1350 \\
0.198 &    6 &      1.92 &      0.992 &      0.981 &       666 &      1030 \\
0.205 &    5 &      2.44 &      1.05 &      0.691 &       558 &      684 \\
0.212 &    4 &      3.03 &      0.775 &      0.801 &       309 &      531 \\
0.215 &    4 &      3.51 &      0.739 &      0.755 &       266 &      460 \\
0.217 &    5 &      3.76 &      0.723 &      0.659 &       265 &      398 \\
0.220 &    5 &      4.27 &      0.692 &      0.572 &       253 &      339 \\
0.221 &    4 &      4.53 &      0.678 &      0.557 &       234 &      319 \\
0.226 &    4 &      5.61 &      0.627 &      0.336 &       288 &      303 \\
0.232 &    3 &      7.01 &      0.572 &      0.719 &       119 &      361 \\
0.236 &    3 &      8.44 &      0.539 &      0.563 &       117 &      272 \\
0.241 &    3 &      9.88 &      0.517 &      0.476 &       116 &      234 \\
0.248 &    3 &      12.8 &      0.485 &      0.346 &       127 &      204 \\
0.261 &    3 &      20.0 &      0.438 &      0.578 &       51.1 &      299 \\
0.266 &    3 &      23.4 &      0.494 &      0.502 &       53.4 &      275 \\
0.278 &    3 &      33.6 &      0.463 &      0.365 &       54.8 &      247 \\
0.284 &    2 &      39.9 &      0.446 &      0.319 &       52.8 &      243 \\
0.287 &    2 &      43.8 &      0.438 &      0.299 &       52.7 &      242 \\
0.290 &    2 &      47.4 &      0.431 &      0.274 &       52.3 &      238 \\
0.293 &    2 &      52.6 &      0.421 &      0.273 &       51.8 &      241 \\
0.306 &    2 &      74.2 &      0.389 &      0.208 &       51.4 &      241 \\
0.317 &    1 &      97.6 &      0.363 &      0.491 &       11.6 &      327 \\
0.322 &    1 &      108 &      0.353 &      0.469 &       11.1 &      326 \\
 \noalign{\smallskip}
 \noalign{\smallskip}
 \end{tabular} 
 \begin{flushleft}
 \end{flushleft}
 \vspace{0.7cm}
 \end{table*}
\clearpage

 \begin{table*}
 \center
 \caption{Properties of selected models. See Table \ref{table:table1} for a description of the parameters.} 
 \begin{tabular}{lllllll}
 \hline  & & & & & & \\ 
 \multicolumn{7}{c}{$Z=0.01$, basic models (no element diffusion nor rotation)}\\ 
 & & & & & & \\ 
 \hline
  Mass ($M_{\odot}$)& $\#$ flashes& $P_{\rm orb,det}$ (d) & $M_{\rm H,det}$ ($10^{-2}\;M_{\odot}$) & $M_{\rm H,T_{\rm eff,max}}$ ($10^{-3}\;M_{\odot}$) & $\Delta t_{\rm proto}$ (Myr) & $t_{\rm cool,L_{-2}}$ (Myr) \\
 \hline
 \hline  
0.176 &     0 &      0.676 &        0.981 &        3.16 &      869 &      0.00 \\
0.192 &    0 &      1.58 &      0.937 &      3.07 &       371 &      2200 \\
0.205 &    0 &      2.76 &      0.944 &      3.08 &       277 &      1120 \\
0.217 &    0 &      4.05 &      1.16 &      3.06 &       188 &      3310 \\
0.222 &    4 &      4.56 &      1.32 &      1.77 &       281 &      2460 \\
0.228 &    3 &      5.05 &      1.51 &      1.66 &       234 &      2290 \\
0.230 &    3 &      5.19 &      0.804 &      1.65 &       125 &      2120 \\
0.237 &    2 &      6.71 &      0.756 &      1.33 &       122 &      1920 \\
0.243 &    2 &      8.26 &      0.721 &      1.60 &       70.3 &      2020 \\
0.248 &    2 &      9.82 &      0.694 &      1.43 &       71.6 &      1930 \\
0.256 &    1 &      12.7 &      0.654 &      1.16 &       89.2 &      1740 \\
0.261 &    1 &      15.2 &      0.627 &      1.55 &       33.9 &      1890 \\
0.269 &    1 &      19.7 &      0.593 &      1.32 &       34.7 &      1740 \\
0.273 &    1 &      21.7 &      0.581 &      1.26 &       35.6 &      1720 \\
0.284 &    1 &      30.5 &      0.539 &      1.03 &       41.2 &      1540 \\
0.297 &    1 &      44.2 &      0.494 &      0.787 &       51.0 &      1320 \\
0.305 &    1 &      55.5 &      0.467 &      0.669 &       59.6 &      1110 \\
0.375 &    0 &      277 &      0.310 &      0.698 &       0.487 &      1010 \\

 \noalign{\smallskip}
 \noalign{\smallskip}
 \end{tabular} 
 \begin{flushleft}
 \end{flushleft}
 \vspace{0.7cm}
 \end{table*}

\clearpage
\begin{table*}
 \center
 \caption{Properties of selected models. See Table \ref{table:table1} for a description of the parameters.} 
 \begin{tabular}{lllllll}
 \hline  & & & & & & \\ 
 \multicolumn{7}{c}{$Z=0.01$, diffusion}\\ 
 & & & & & & \\ 
 \hline
 Mass ($M_{\odot}$)& $\#$ flashes& $P_{\rm orb,det}$ (d) & $M_{\rm H,det}$ ($10^{-2}\;M_{\odot}$) & $M_{\rm H,T_{\rm eff,max}}$ ($10^{-3}\;M_{\odot}$) & $\Delta t_{\rm proto}$ (Myr) & $t_{\rm cool,L_{-2}}$ (Myr) \\
 \hline
 \hline  
0.165 &      0 &      0.340 &        0.904 &        4.45 &      1580 &      0.00 \\
0.170 &    7 &      0.458 &      0.946 &      0.807 &       2380 &      2750 \\
0.171 &    5 &      0.501 &      0.950 &      0.967 &       2320 &      2780 \\
0.175 &    5 &      0.651 &      1.02 &      0.980 &       1910 &      2350 \\
0.181 &    7 &      0.937 &      1.05 &      0.589 &       1500 &      1590 \\
0.203 &    6 &      2.97 &      0.927 &      0.804 &       487 &      731 \\
0.212 &    5 &      4.26 &      1.01 &      0.707 &       390 &      572 \\
0.223 &    5 &      5.53 &      1.30 &      0.448 &       405 &      447 \\
0.232 &    4 &      6.74 &      0.784 &      0.593 &       183 &      339 \\
0.238 &    4 &      8.33 &      0.745 &      0.411 &       201 &      268 \\
0.240 &    2 &      8.99 &      0.729 &      1.23 &       130 &      1810 \\
0.243 &    3 &      9.98 &      0.712 &      0.808 &       94.8 &      415 \\
0.248 &    3 &      11.6 &      0.685 &      0.654 &       95.9 &      335 \\
0.252 &    3 &      13.2 &      0.664 &      0.561 &       97.2 &      293 \\
0.255 &    4 &      14.7 &      0.646 &      0.475 &       102 &      261 \\
0.260 &    3 &      17.4 &      0.617 &      0.371 &       114 &      232 \\
0.265 &    3 &      19.9 &      0.596 &      0.299 &       127 &      219 \\
0.271 &    3 &      24.2 &      0.566 &      0.658 &       47.1 &      342 \\
0.274 &    2 &      26.2 &      0.555 &      0.572 &       47.6 &      311 \\
0.279 &    3 &      29.9 &      0.536 &      0.534 &       48.5 &      301 \\
0.283 &    2 &      33.2 &      0.522 &      0.474 &       49.2 &      283 \\
0.286 &    2 &      36.4 &      0.510 &      0.425 &       49.6 &      268 \\
0.291 &    2 &      42.0 &      0.491 &      0.377 &       50.3 &      260 \\
 \noalign{\smallskip}
 \noalign{\smallskip}
 \end{tabular} 
 \begin{flushleft}
 \end{flushleft}
 \vspace{0.7cm}
 \end{table*}

\begin{table*}
 \center
 \caption{Properties of selected models. See Table \ref{table:table1} for a description of the parameters.} 
 \begin{tabular}{lllllll}
 \hline  & & & & & & \\ 
 \multicolumn{7}{c}{$Z=0.01$, diffusion+rotation}\\ 
 & & & & & & \\ 
 \hline
  Mass ($M_{\odot}$)& $\#$ flashes& $P_{\rm orb,det}$ (d) & $M_{\rm H,det}$ ($10^{-2}\;M_{\odot}$) & $M_{\rm H,T_{\rm eff,max}}$ ($10^{-3}\;M_{\odot}$) & $\Delta t_{\rm proto}$ (Myr) & $t_{\rm cool,L_{-2}}$ (Myr) \\
 \hline
 \hline  
0.182 &    7 &      1.01 &      1.04 &      0.873 &       1390 &      1740 \\
0.183 &    6 &      1.09 &      1.04 &      0.898 &       1270 &      1620 \\
0.187 &    7 &      1.34 &      1.04 &      0.946 &       1070 &      1460 \\
0.192 &    7 &      1.72 &      1.02 &      0.957 &       823 &      1210 \\
0.195 &    7 &      1.92 &      1.02 &      0.790 &       777 &      1020 \\
0.197 &    6 &      2.14 &      0.983 &      0.995 &       615 &      1020 \\
0.206 &    5 &      3.32 &      0.958 &      0.689 &       491 &      652 \\
0.216 &    5 &      4.59 &      1.09 &      0.824 &       339 &      618 \\
0.226 &    4 &      5.79 &      1.41 &      0.870 &       264 &      596 \\
0.234 &    4 &      7.05 &      0.775 &      0.543 &       187 &      317 \\
0.239 &    4 &      8.58 &      0.740 &      0.399 &       199 &      262 \\
0.244 &    3 &      10.2 &      0.709 &      0.723 &       102 &      378 \\
0.248 &    3 &      11.7 &      0.684 &      0.660 &       95.1 &      342 \\
0.252 &    4 &      13.2 &      0.663 &      0.565 &       98.9 &      293 \\
0.258 &    3 &      16.0 &      0.631 &      0.401 &       111 &      240 \\
0.262 &    3 &      18.5 &      0.608 &      0.326 &       121 &      222 \\
0.265 &    3 &      19.9 &      0.596 &      0.299 &       127 &      219 \\
0.275 &    2 &      26.8 &      0.551 &      0.579 &       48.1 &      314 \\
0.279 &    2 &      30.4 &      0.533 &      0.501 &       48.9 &      289 \\
0.301 &    2 &      54.3 &      0.460 &      0.290 &       52.3 &      251 \\
0.324 &    1 &      90.6 &      0.401 &      0.556 &       12.3 &      368 \\
 \noalign{\smallskip}
 \noalign{\smallskip}
 \end{tabular} 
 \begin{flushleft}
 \end{flushleft}
 \vspace{0.7cm}
 \end{table*}

\clearpage
\begin{table*}
 \center
 \caption{Properties of selected models. See Table \ref{table:table1} for a description of the parameters.} 
 \begin{tabular}{lllllll}
 \hline  & & & & & & \\ 
 \multicolumn{7}{c}{$Z=0.001$, basic models (no element diffusion nor rotation)}\\ 
 & & & & & & \\ 
 \hline
  Mass ($M_{\odot}$)& $\#$ flashes& $P_{\rm orb,det}$ (d)& $M_{\rm H,det}$ ($10^{-2}\;M_{\odot}$) & $M_{\rm H,T_{\rm eff,max}}$ ($10^{-3}\;M_{\odot}$) & $\Delta t_{\rm proto}$ (Myr) & $t_{\rm cool,L_{-2}}$ (Myr) \\
 \hline
 \hline  
0.182 &    0 &      0.471 &      1.48 &      3.74 &       1100 &      2640 \\
0.190 &    0 &      0.674 &      1.55 &      3.71 &       816 &      2860 \\
0.207 &    0 &      1.45 &      1.58 &      3.64 &       387 &      3550 \\
0.211 &    0 &      1.75 &      1.56 &      3.62 &       314 &      3410 \\
0.228 &    0 &      3.49 &      1.44 &      3.44 &       142 &      2760 \\
0.234 &    0 &      4.62 &      1.37 &      3.32 &       102 &      2610 \\
0.238 &    0 &      5.38 &      1.32 &      3.25 &       85.5 &      2470 \\
0.246 &    0 &      7.22 &      1.23 &      3.10 &       60.2 &      2310 \\
0.252 &    3 &      8.90 &      1.17 &      1.85 &       88.9 &      1760 \\
0.256 &    4 &      10.4 &      1.10 &      1.87 &       72.2 &      1720 \\
0.260 &    4 &      11.8 &      1.05 &      1.63 &       73.2 &      1630 \\
0.263 &    3 &      13.0 &      1.02 &      1.79 &       60.6 &      1690 \\
0.266 &    3 &      14.0 &      0.999 &      1.97 &       61.7 &      1780 \\
0.274 &    3 &      17.5 &      1.01 &      1.48 &       61.8 &      1540 \\
0.282 &    3 &      20.0 &      1.16 &      1.27 &       65.5 &      1430 \\
0.289 &    2 &      22.0 &      1.42 &      1.55 &       43.1 &      1510 \\
0.302 &    2 &      26.9 &      0.861 &      1.11 &       45.5 &      1310 \\
0.309 &    1 &      31.9 &      0.815 &      1.56 &       15.6 &      1450 \\
0.328 &    1 &      49.1 &      0.708 &      1.16 &       16.6 &      1270 \\
0.340 &    1 &      63.7 &      0.648 &      0.950 &       18.9 &      1150 \\
0.350 &    1 &      76.9 &      0.607 &      0.810 &       21.4 &      1090 \\
0.423 &    0 &      258 &      0.385 &      0.781 &       0.406 &      875 \\
 \noalign{\smallskip}
 \noalign{\smallskip}
 \end{tabular} 
 \begin{flushleft}
 \end{flushleft}
 \vspace{0.7cm}
 \end{table*}

\clearpage
\begin{table*}
 \center
 \caption{Properties of selected models. See Table \ref{table:table1} for a description of the parameters.} 
 \begin{tabular}{lllllll}
 \hline  & & & & & & \\ 
 \multicolumn{7}{c}{$Z=0.001$, diffusion}\\ 
 & & & & & & \\ 
 \hline
  Mass ($M_{\odot}$)& $\#$ flashes& $P_{\rm orb,det}$ (d) & $M_{\rm H,det}$ ($10^{-2}\;M_{\odot}$) & $M_{\rm H,T_{\rm eff,max}}$ ($10^{-3}\;M_{\odot}$) & $\Delta t_{\rm proto}$ (Myr) & $t_{\rm cool,L_{-2}}$ (Myr) \\
 \hline
 \hline  
0.180 &      0 &      0.466 &        1.50 &        5.00 &      1280 &      0.00 \\
0.183 &      5 &      0.582 &        1.58 &        4.86 &      1160 &      0.00 \\
0.190 &    0 &      0.800 &      1.64 &      4.63 &       886 &      5900 \\
0.196 &    0 &      1.05 &      1.67 &      4.44 &       692 &      5830 \\
0.217 &    1 &      2.65 &      1.60 &      3.87 &       261 &      3920 \\
0.230 &    4 &      4.59 &      1.46 &      1.21 &       289 &      805 \\
0.239 &    6 &      6.61 &      1.35 &      0.840 &       227 &      589 \\
0.245 &    5 &      8.54 &      1.27 &      0.958 &       155 &      580 \\
0.250 &    2 &      10.3 &      1.20 &      0.674 &       168 &      451 \\
0.254 &    2 &      11.9 &      1.15 &      0.651 &       153 &      429 \\
0.264 &    2 &      16.7 &      1.01 &      0.668 &       122 &      412 \\
0.268 &    1 &      18.5 &      0.985 &      0.518 &       149 &      360 \\
0.271 &    1 &      19.9 &      0.986 &      0.462 &       153 &      341 \\
0.280 &    3 &      23.4 &      1.11 &      0.756 &       77.6 &      438 \\
0.314 &    1 &      41.4 &      0.785 &      0.339 &       76.9 &      282 \\
0.322 &    3 &      49.7 &      0.734 &      0.676 &       28.1 &      452 \\
 \noalign{\smallskip}
 \noalign{\smallskip}
 \end{tabular} 
 \begin{flushleft}
 \end{flushleft}
 \vspace{0.7cm}
 \end{table*}

\begin{table*}
 \center
 \caption{Properties of selected models. See Table \ref{table:table1} for a description of the parameters.} 
 \begin{tabular}{lllllll}
 \hline  & & & & & & \\ 
 \multicolumn{7}{c}{$Z=0.001$, diffusion+rotation}\\ 
 & & & & & & \\
 \hline
 Mass ($M_{\odot}$)& $\#$ flashes& $P_{\rm orb,det}$ (d)& $M_{\rm H,det}$ ($10^{-2}\;M_{\odot}$) & $M_{\rm H,T_{\rm eff,max}}$ ($10^{-3}\;M_{\odot}$) & $\Delta t_{\rm proto}$ (Myr) & $t_{\rm cool,L_{-2}}$ (Myr) \\
 \hline
 \hline  
0.179 &      0 &      0.477 &        1.54 &        5.30 &      1310 &      0.00 \\
0.183 &      0 &      0.572 &        1.59 &        5.18 &      1140 &      0.00 \\
0.190 &    0 &      0.783 &      1.65 &      4.86 &       891 &      2300 \\
0.196 &    0 &      1.03 &      1.68 &      4.62 &       702 &      2230 \\
0.216 &    0 &      2.59 &      1.59 &      4.02 &       270 &      4180 \\
0.228 &    0 &      4.55 &      1.43 &      3.74 &       144 &      2800 \\
0.237 &      4 &      6.58 &        1.32 &        1.18 &      233 &      0.00 \\
0.244 &    1 &      8.49 &      1.23 &      0.646 &       246 &      496 \\
0.250 &    5 &      10.2 &      1.18 &      0.986 &       138 &      582 \\
0.254 &    5 &      11.7 &      1.16 &      0.803 &       132 &      487 \\
0.260 &    5 &      14.4 &      1.07 &      0.808 &       109 &      479 \\
0.275 &    1 &      21.1 &      1.03 &      0.443 &       144 &      326 \\
0.287 &    1 &      25.0 &      1.35 &      0.489 &       102 &      332 \\
0.296 &    2 &      27.3 &      0.910 &      0.725 &       50.4 &      435 \\
0.303 &    1 &      32.3 &      0.858 &      0.558 &       58.0 &      351 \\
0.314 &    1 &      41.5 &      0.785 &      0.341 &       78.9 &      284 \\
0.322 &    1 &      49.7 &      0.734 &      0.644 &       28.3 &      428 \\

 \noalign{\smallskip}
 \noalign{\smallskip}
 \end{tabular} 
 \begin{flushleft}
 \end{flushleft}
 \vspace{0.7cm}
 \end{table*}

\clearpage
\begin{table*}
 \center
 \caption{Properties of selected models. See Table \ref{table:table1} for a description of the parameters.} 
 \begin{tabular}{lllllll}
 \hline  & & & & & & \\ 
 \multicolumn{7}{c}{$Z=0.0002$, basic models (no element diffusion nor rotation)}\\ 
 & & & & & & \\ 
 \hline
 Mass ($M_{\odot}$)& $\#$ flashes& $P_{\rm orb,det}$ (d) & $M_{\rm H,det}$ ($10^{-2}\;M_{\odot}$) & $M_{\rm H,T_{\rm eff,max}}$ ($10^{-3}\;M_{\odot}$) & $\Delta t_{\rm proto}$ (Myr) & $t_{\rm cool,L_{-2}}$ (Myr) \\
 \hline
 \hline  
0.214 &    0 &      0.900 &      2.10 &      3.78 &       533 &      3550 \\
0.227 &    0 &      1.61 &      2.10 &      3.74 &       318 &      2980 \\
0.247 &    0 &      3.86 &      1.90 &      3.58 &       131 &      2350 \\
0.260 &    0 &      6.21 &      1.74 &      3.42 &       76.6 &      2110 \\
0.268 &    0 &      8.30 &      1.63 &      3.28 &       54.9 &      1940 \\
0.274 &    0 &      10.1 &      1.56 &      3.17 &       43.9 &      1870 \\
0.278 &    0 &      11.6 &      1.51 &      3.10 &       37.4 &      1810 \\
0.282 &    0 &      12.8 &      1.47 &      3.03 &       33.2 &      1780 \\
0.284 &    1 &      13.9 &      1.43 &      2.44 &       41.0 &      1750 \\
0.287 &    3 &      14.9 &      1.40 &      1.98 &       47.6 &      1480 \\
0.289 &    4 &      15.8 &      1.38 &      1.84 &       47.0 &      1410 \\
0.290 &    4 &      16.6 &      1.36 &      1.83 &       45.6 &      1410 \\
0.292 &    4 &      17.3 &      1.34 &      1.73 &       44.7 &      1370 \\
0.293 &    3 &      18.0 &      1.32 &      1.96 &       36.1 &      1420 \\
0.295 &    3 &      18.7 &      1.31 &      1.82 &       37.0 &      1370 \\
0.297 &    3 &      19.9 &      1.28 &      1.60 &       39.2 &      1310 \\
0.302 &    4 &      23.1 &      1.20 &      1.50 &       36.9 &      1270 \\
0.305 &    3 &      25.0 &      1.16 &      1.90 &       25.6 &      1380 \\
0.308 &    4 &      26.8 &      1.12 &      1.61 &       31.8 &      1310 \\
0.313 &    3 &      30.2 &      1.07 &      1.65 &       27.2 &      1300 \\
0.317 &    2 &      33.4 &      1.04 &      1.85 &       20.0 &      1340 \\
0.320 &    2 &      34.9 &      1.05 &      1.74 &       20.4 &      1300 \\
0.322 &    2 &      36.4 &      1.06 &      1.65 &       21.0 &      1280 \\
0.327 &    2 &      39.1 &      1.12 &      1.47 &       23.2 &      1210 \\
0.329 &    2 &      40.4 &      1.18 &      1.38 &       25.0 &      1180 \\
0.336 &    2 &      43.1 &      1.39 &      1.13 &       33.4 &      1100 \\
0.340 &    1 &      44.6 &      1.55 &      1.68 &       16.0 &      1250 \\
0.342 &    1 &      45.5 &      1.06 &      1.62 &       11.6 &      1220 \\
0.356 &    1 &      57.5 &      0.901 &      1.32 &       10.7 &      1110 \\
0.441 &    0 &      232 &      0.514 &      0.940 &       0.552 &      829 \\
 \noalign{\smallskip}
 \end{tabular} 
 \begin{flushleft}
 \end{flushleft}
 \vspace{0.7cm}
 \end{table*}

\clearpage
\begin{table*}
 \center
 \caption{Properties of selected models. See Table \ref{table:table1} for a description of the parameters.} 
 \begin{tabular}{lllllll}
 \hline  & & & & & & \\ 
 \multicolumn{7}{c}{$Z=0.0002$, diffusion}\\ 
 & & & & & & \\ 
 \hline
 Mass ($M_{\odot}$)& $\#$ flashes& $P_{\rm orb,det}$ (d) & $M_{\rm H,det}$ ($10^{-2}\;M_{\odot}$) & $M_{\rm H,T_{\rm eff,max}}$ ($10^{-3}\;M_{\odot}$) & $\Delta t_{\rm proto}$ (Myr) & $t_{\rm cool,L_{-2}}$ (Myr) \\
 \hline
 \hline  
0.226 &    0 &      1.78 &      2.23 &      4.18 &       374 &      4060 \\
0.234 &    0 &      2.54 &      2.17 &      4.01 &       265 &      3670 \\
0.244 &    0 &      4.00 &      2.03 &      3.81 &       166 &      3240 \\
0.246 &    0 &      4.40 &      2.00 &      3.77 &       149 &      3130 \\
0.255 &    0 &      6.38 &      1.86 &      3.60 &       98.3 &      2830 \\
0.263 &    0 &      8.56 &      1.74 &      3.43 &       69.8 &      2640 \\
0.269 &    5 &      10.5 &      1.66 &      1.19 &       122 &      735 \\
0.273 &    0 &      12.1 &      1.60 &      3.22 &       46.4 &      2420 \\
0.277 &    4 &      13.5 &      1.55 &      0.919 &       110 &      569 \\
0.282 &    4 &      15.9 &      1.48 &      0.944 &       89.1 &      573 \\
0.286 &    5 &      17.8 &      1.43 &      0.952 &       76.1 &      574 \\
0.289 &    5 &      19.4 &      1.39 &      0.678 &       100 &      441 \\
0.290 &    5 &      20.1 &      1.38 &      0.644 &       103 &      420 \\
0.291 &    4 &      20.8 &      1.36 &      1.03 &       59.3 &      639 \\
0.297 &    4 &      24.3 &      1.28 &      1.04 &       48.8 &      664 \\
0.300 &    4 &      26.3 &      1.24 &      0.899 &       53.4 &      570 \\
0.311 &    5 &      34.8 &      1.08 &      0.657 &       66.5 &      412 \\
 \noalign{\smallskip}
 \noalign{\smallskip}
 \end{tabular} 
 \begin{flushleft}
 \end{flushleft}
 \vspace{0.7cm}
 \label{table:z_00002_diffusion}
 \end{table*}

\begin{table*}
 \center
 \caption{Properties of selected models. See Table \ref{table:table1} for a description of the parameters.} 
 \begin{tabular}{lllllll}
 \hline  & & & & & & \\ 
 \multicolumn{7}{c}{$Z=0.0002$, diffusion+rotation}\\ 
  & & & & & & \\ 
 \hline
 Mass ($M_{\odot}$)& $\#$ flashes& $P_{\rm orb,det}$ (d) & $M_{\rm H,det}$ ($10^{-2}\;M_{\odot}$) & $M_{\rm H,T_{\rm eff,max}}$ ($10^{-3}\;M_{\odot}$) & $\Delta t_{\rm proto}$ (Myr) & $t_{\rm cool,L_{-2}}$ (Myr) \\
 \hline
 \hline  
0.212 &      0 &      0.902 &        2.17 &        4.79 &      621 &      0.00 \\
0.236 &    0 &      2.90 &      2.13 &      4.11 &       229 &      3290 \\
0.250 &    0 &      5.32 &      1.92 &      3.77 &       121 &      3100 \\
0.259 &    0 &      7.67 &      1.76 &      3.61 &       80.9 &      2830 \\
0.266 &      0 &      9.76 &        1.65 &        3.51 &      61.8 &      0.00 \\
0.270 &    0 &      11.5 &      1.57 &      3.46 &       51.0 &      2590 \\
0.274 &    0 &      13.1 &      1.52 &      3.40 &       44.5 &      2560 \\
0.277 &    3 &      14.4 &      1.47 &      1.08 &       115 &      688 \\
0.280 &    5 &      15.6 &      1.43 &      1.19 &       89.2 &      779 \\
0.282 &    5 &      16.6 &      1.40 &      1.10 &       89.1 &      692 \\
0.284 &    4 &      17.5 &      1.38 &      0.875 &       103 &      542 \\
0.286 &    5 &      18.4 &      1.36 &      0.892 &       95.9 &      556 \\
0.287 &    5 &      19.1 &      1.34 &      0.836 &       94.3 &      519 \\
0.288 &    5 &      19.9 &      1.32 &      0.893 &       84.0 &      558 \\
0.290 &    5 &      20.6 &      1.32 &      0.872 &       82.0 &      538 \\
0.291 &    5 &      21.2 &      1.31 &      0.897 &       75.2 &      552 \\
0.293 &    5 &      21.8 &      1.30 &      0.868 &       74.9 &      541 \\

 \noalign{\smallskip}
 \noalign{\smallskip}
 \end{tabular} 
 \begin{flushleft}
 \end{flushleft}
 \vspace{0.7cm}
 \label{table:z_00002_rotation}
 \end{table*}

\end{document}